\documentclass[MSNbibl,nameyear,rotating,dvips]{arxstspdf}
\usepackage{multirow}
\usepackage{dcolumn}
\usepackage{flushend}
\usepackage{stfloats}
\usepackage{graphicx}

\volume{28}
\issue{2}
\pubyear{2013}
\firstpage{168}
\lastpage{188}
\doi{10.1214/13-STS418} 

\makeatletter

\newcolumntype{d}[1]{D{.}{.}{#1}}

\newcommand{\rright}{\right}
\newcommand{\lleft}{\left}

\newtheorem{lemma}{Lemma}

\setattribute{abstract}{width}{360pt}
\setattribute{keyword}{width}{360pt}
\makeatother

\begin{document}
\begin{frontmatter}

\title{Variational Inference for Generalized Linear Mixed Models
Using Partially Noncentered Parametrizations}
\runtitle{Variational Inference for GLMMs Using Partially Noncentered
Parametrizations}

\begin{aug}
\author[a]{\fnms{Linda S. L.} \snm{Tan}\corref{}\ead[label=e1]{g0900760@nus.edu.sg}}
\and
\author[a]{\fnms{David J.} \snm{Nott}\ead[label=e2]{standj@nus.edu.sg}}
\runauthor{L. S. L. Tan and D. J. Nott}

\affiliation{National University of Singapore}

\address[a]{Linda S. L. Tan is a Ph.D. student and David J. Nott is
Associate Professor,
Department of Statistics and
Applied Probability, National University of Singapore, Singapore
117546, Singapore \printead{e1,e2}.}

\end{aug}

%
\begin{abstract}
The effects of different parametrizations on the convergence of
Bayesian computational algorithms for hierarchical models are well
explored. Techniques such as centering, noncentering and partial
noncentering can be used to accelerate convergence in MCMC and EM
algorithms but are still not well studied for variational Bayes (VB)
methods. As a fast deterministic approach to posterior approximation,
VB is attracting increasing interest due to its suitability for large
high-dimensional data. Use of different parametrizations for VB has not
only computational but also statistical implications, as different
parametrizations are associated with different factorized posterior
approximations. We examine the use of partially noncentered
parametrizations in VB for generalized linear mixed models (GLMMs). Our
paper makes four contributions. First, we show how to implement an
algorithm called nonconjugate variational message passing for GLMMs.
Second, we show that the partially noncentered parametrization can
adapt to the quantity of information in the data and determine a
parametrization close to optimal. Third, we show that partial
noncentering can accelerate convergence and produce more accurate
posterior approximations than centering or noncentering. Finally, we
demonstrate how the variational lower bound, produced as part of the
computation, can be useful for model selection.
\end{abstract}

%

\end{frontmatter}

\section{Introduction}\label{sec1}
The convergence of Markov chain Monte Carlo (MCMC) algorithms depends
greatly on the choice of parametrization and simple reparametrizations
can often give improved convergence. Here we investigate the use of
centered, noncentered and partially noncentered parametrizations of
hierarchical models in the context of variational Bayes (VB)
(\cite{Att99}). As a fast deterministic approach to approximation of the
posterior distribution in Bayesian inference, VB is attracting
increasing interest due to its suitability for large high-dimensional
data (see, e.g., \cite{BraMcA10}; \cite{Hofetal}). VB methods approximate the
intractable posterior by a factorized distribution which can be
represented by a directed graph and optimization of the factorized
\mbox{variational} posterior can be decomposed into local computations that
involve only neighboring nodes. Variational message passing\break
(\cite{WinBis05}) is an algorithmic implementation of VB that can be
applied to a general class of conjugate-exponential models
(\cite{Att}; \cite{GhaBea}). \citet{KnoMin} proposed an algorithm called a
nonconjugate variational message passing to extend variational message
passing to nonconjugate models.

We examine the use of partially noncentered pa\-rametrization in VB for
generalized linear mixed models (GLMMs). Our paper makes four
contributions. First, we show how to implement nonconjugate variational
message passing for GLMMs. Second, we show that the partially
noncentered parame\-trization is able to adapt to the quantity of
information in the data so that it is not necessary to make a choice in
advance between centering and noncentering with the data deciding the
optimal parametrization. Third, we show that in addition to
accelerating convergence, partial noncentering is a good strategy
statistically for VB in terms of producing more accurate approximations
to the posterior than either centering or noncentering. Finally, we
demonstrate how the variational lower bound, which is produced as part
of the computation, can be useful for model selection.

GLMMs extend generalized linear models by the inclusion of random
effects to account for correlation of observations in grouped data and
are of wide applicability. Estimation of GLMMs using maximum likelihood
is challenging, as the integral over random effects is intractable.
Methods involving numerical quadrature or MCMC to approximate these
integrals are computationally intensive. Various approximate methods
such as penalized quasi-likelihood (\cite{BreCla93}), Laplace
approximation and its extension (\cite{RauYanYos00}) and
Gaussian variational approximation\break (\cite{OrmWan12}) have been
developed. \citet{FonRueWak10} considered a Bayes\-ian approach
using integrated nested Laplace approximations. We show how to fit
GLMMs using nonconjugate variational message passing, focusing on
Poisson and logistic mixed models and their applications in
longitudinal data analysis.

The literature on parametrization of hierarchical models including
partial noncentering techniques for accelerating MCMC algorithms is
inspired by earlier similar work for the expectation maximization (EM)
algorithm (see Meng and van Dyk, \citeyear{Menvan97}, \citeyear{Menvan99};
\cite{LiuWu99}). Gelfand, Sahu and Carlin
(\citeyear{GelSahCar95}, \citeyear{GelSahCar96}) proposed hierarchical centering for normal
linear mixed models and GLMMs to improve the slow mixing in MCMC
algorithms due to high correlations between model parameters.
Papaspiliopoulos, Roberts and Sk{\"o}ld (\citeyear{PapRobSko03},
\citeyear{PapRobSko07}) demonstrated that centering
and noncentering play complementary\break roles in boosting MCMC efficiency
and neither are uniformly effective. They considered the partially
noncentered parametrization which is data dependent and lies on the
continuum between the centered and noncentered parametrizations.
Extending this idea, \citet{ChrRobSko06} devised
re\-parametrization techniques to improve performance for Hastings-within
Gibbs algorithms for spatial\break GLMMs. Yu and Meng (\citeyear{YuMen11}) introduced a
strategy for boosting MCMC efficiency via interweaving the centered and
noncentered parametrizations to reduce dependence between draws.
Parameter-expanded VB methods were proposed by \citet{QiJaa}
to reduce coupling in updates and speed up VB.

The idea of partial noncentering is to introduce a tuning parameter via
reparametrization of the model and then seek its optimal value for
fastest convergence. For the normal hierarchical model,
Papaspilio\-poulos, Roberts and Sk{\"o}ld (\citeyear{PapRobSko03}) showed that the partially
noncentered parametrization has convergence properties superior to that
of the centered and noncentered parametrizations for the Gibbs sampler.
As the rate of convergence of an algorithm based on VB is equal to that
of the corresponding Gibbs sampler when the target distribution is
Gaussian (\cite{TanNot}), partial noncentering will similarly
outperform centering and noncentering in the context of VB for the
normal hierarchical model. This provides motivation to consider partial
noncentering in the VB context. We illustrate this idea with the
following example.

{\renewcommand{\figurename}{Algorithm}
%
\begin{figure*}
\begin{tabular}{l}
\rule{342pt}{0.5pt}\\
Initialize $\mu_{\tilde{\alpha}_i}^q$ and $\Sigma_{\tilde{\alpha}_i}^q$
for $i=1,\ldots,n$.\\
Cycle: \\
\hphantom{Cycle:}$\Sigma_\beta^q \leftarrow [ \sum_{i=1}^n  \{
(I-W_i)^TD^{-1}(I-W_i)+\frac{1}{\sigma^2}W_i^T X_i^TX_iW_i  \}
]^{-1} $ \\
\hphantom{Cycle:}$\mu_\beta^q \leftarrow\Sigma_\beta^q \sum_{i=1}^n
[ \frac{1}{\sigma^2} W_i^T X_i^T y_i +\{D^{-1}(I-W_i)-\frac
{1}{\sigma^2}X_i^TX_iW_i\}^T\mu_{\tilde{\alpha}_i}^q  ]$ \\
\hphantom{Cycle:}For $i=1,\ldots,n$,\\
\hphantom{Cycle:For }$\Sigma_{\tilde{\alpha}_i}^q \leftarrow
(D^{-1}+ \frac{1}{\sigma^2}X_i^TX_i  )^{-1}$ \\
\hphantom{Cycle:For }$\mu_{\tilde{\alpha}_i}^q \leftarrow\Sigma_{\tilde
{\alpha}_i}^q [ \frac{1}{\sigma^2}X_i^Ty_i+\{D^{-1}(I-W_i)-\frac
{1}{\sigma^2}X_i^TX_iW_i\}\mu_\beta^q  ]$ \\
until convergence.\\[-6pt]
\rule{342pt}{0.5pt}
\end{tabular}
\caption{Iterative scheme for obtaining variational parameters in
linear mixed model.}\label{algor1}
\end{figure*}}

\subsection{Motivating Example: Linear Mixed Model}\label{sec1.1}
Consider the linear mixed model
%
\begin{eqnarray}\label{linearmixedmodel}
y_i = X_i\beta+ X_i u_i +
\varepsilon_i,\nonumber\\[-8pt]\\[-8pt]
&&\eqntext{\varepsilon_i \sim N\bigl(0,
\sigma^2I\bigr), i=1,\ldots,n,}
\end{eqnarray}
where $y_i$ is a vector of length $n_i$, $\beta$ is a vector of length
$r$ of fixed effects, $X_i$ is a $n_i \times r$ matrix of covariates
and $u_i$ is a vector of length $r$ of random effects independently
distributed as $N(0,D)$. For simplicity,\vspace*{1pt} we specify a constant prior on
$\beta$ and assume $\sigma^2$ and $D$ are known. Let
\[
\alpha_i = \beta+u_i \quad\mbox{and}\quad \tilde{
\alpha}_i=\alpha _i-W_i\beta,\quad i=1,\ldots,n,
\]
where $W_i$ is an $r \times r$ tuning matrix to be specified. $W_i=0$
corresponds to the centered and $W_i=I$ to the noncentered
parametrization. For each $i=1,\ldots,n$,
\[
y_i = X_iW_i\beta+ X_i
\tilde{\alpha}_i + \varepsilon_i
\]
and
\[
\tilde{
\alpha}_i \sim N \bigl((I-W_i)\beta,D \bigr).
\]
This is the partially noncentered parametrization and the set of
unknown parameters is $\theta=\{\beta,\tilde{\alpha}\}$, where $\tilde
{\alpha}=[\tilde{\alpha}_1^T,\ldots,\tilde{\alpha}_n^T]^T$. Let
$y=[y_1,\ldots,y_n]^T$ denote the observed data. Of interest is the
posterior distribution of $\theta$, $p(\theta|y)$.

Suppose $p(\theta|y)$ is not analytically tractable. In the variational
approach, we approximate $p(\theta|y)$ by a $q(\theta)$ for which
inference is more tractable and $q(\theta)$ is chosen to minimize the
Kullback--Leibler divergence between $q(\theta)$ and $p(\theta|y)$ given by
\begin{eqnarray*}
\int q(\theta)\log\frac{q(\theta)}{p(\theta|y)} \,\mathrm{d}\theta &=& \int q(\theta) \log
\frac{q(\theta)}{p(y,\theta)} \,\mathrm{d}\theta\\
&&{} + \log p(y),
\end{eqnarray*}
where $p(y)$ is the marginal likelihood $p(y) = \int p(y|\allowbreak\theta) p(\theta
) \,\mathrm{d}\theta$.
Since the Kullback--Leibler divergence is nonnegative,
%
\begin{eqnarray}\label{lowerbound}
\log p(y) &\geq&\int\log\frac{p(y,\theta)}{q(\theta)} q(\theta) \,\mathrm{d}\theta
\nonumber
\\
&=&E_q\bigl\{\log p(y,\theta)\bigr\}-E_q\bigl\{\log q(
\theta)\bigr\}
\\
&=&\mathcal{L},\nonumber
\end{eqnarray}
where $\mathcal{L}$ is a lower bound on the log marginal likelihood.
Maximization of $\mathcal{L}$ is equivalent to minimization of the
Kullback--Leibler divergence between $q(\theta)$ and $p(\theta|y)$. In
VB, $q(\theta)$ is assumed to be of a factorized form, say, $q(\theta
)=\prod_{i=1}^m q_i(\theta_i)$ for some partition $\{\theta_1,\ldots,\theta
_m\}$ of $\theta$. Maximization of $\mathcal{L}$ over each of
$q_1,\ldots,q_m$ lead to optimal densities satisfying
$q_i(\theta_i) \propto\exp\{E_{-\theta_i}\log p(y,\theta)\}$,
$i=1,\ldots,m$,
where $E_{-\theta_i}$ denotes expectation with respect to the density\break
$\prod_{j \neq i} q_j(\theta_j)$. See \citet{OrmWan10} for an
explanation of variational approximation methods very accessible to
statisticians.

If we apply VB to (\ref{linearmixedmodel}) and approximate the
posterior $p(\theta|y)$ with $q(\theta)=q(\beta)q(\tilde{\alpha})$, the
optimal densities can be derived to be
$q(\beta)=N(\mu_\beta^q,\Sigma_\beta^q)$ and
$q(\tilde{\alpha})=\prod_{i=1}^n q(\tilde{\alpha}_i)$, where
$q(\tilde{\alpha}_i)=N(\mu_{\tilde{\alpha}_i}^q,\Sigma_{\tilde{\alpha}_i}^q)$.
The expressions for the variational parameters $\mu_\beta^q$,
$\Sigma_\beta^q$ and $\mu_{\tilde{\alpha}_i}^q$,
$\Sigma_{\tilde{\alpha}_i}^q$, $i=1,\ldots,m$, are,\vspace*{1pt} however, dependent on
each other and can be computed by an iterative scheme such as that
given in Algorithm~\ref{algor1}.

Observe that Algorithm~\ref{algor1} converges in one iteration if
$D^{-1}(I-W_i)=\frac{1}{\sigma^2}X_i^TX_iW_i$ for each $i$, that is, if
%
\begin{eqnarray}\label{Wi}
W_i=\biggl(\frac{1}{\sigma^2}X_i^T
X_i +D^{-1}\biggr)^{-1}D^{-1}\nonumber\\[-8pt]\\[-8pt]
&&\eqntext{\mbox{for } i=1,\ldots,n.}
\end{eqnarray}
For this specification of the tuning parameters, partial noncentering
gives more rapid convergence than centering or noncentering. Moreover,
it can be shown that the true posteriors are recovered in this
partially noncentered parametrization so that a better fit is achieved
than in the centered or noncentered parametrizations. This example
suggests that with careful tuning of $W_i$, $i=1,\ldots,n$, the partially
noncentered parametrization can potentially outperform the centered and
noncentered parametrizations in the VB context.

The rest of the paper is organized as follows. Section~\ref{sec2} specifies the
GLMM and priors used. Section~\ref{sec3} describes the partially noncentered
parametrization for GLMMs. Section~\ref{sec4} describes the nonconjugate
variational message passing algorithm for fitting\break GLMMs. Section
\ref{sec5}
discusses briefly the use of the variational\vadjust{\goodbreak} lower bound for model
selection and Section~\ref{sec6} considers examples including real and simulated
data. Section~\ref{sec7} concludes.

\section{The Generalized Linear Mixed Model}\label{sec2}
Consider clustered data where $y_{ij}$ denotes the $j$th response from
cluster $i$, $i=1,\ldots, n$, $j=1,\ldots,n_i$. Conditional on the
$r$-dimensional random effects $u_i$ drawn independently from $N(0,
D)$, $y_{ij}$ is independently distributed from some exponential family
distribution with density
%
\begin{equation}\label{GLMM}\quad
f(y_{ij}|u_i) = \exp \biggl\{\frac{y_{ij}\zeta_{ij}-b(\zeta_{ij})}{a(\phi
)}+c(y_{ij},
\phi) \biggr\},
\end{equation}
where $\zeta_{ij}$ is the canonical parameter, $\phi$ is the dispersion
parameter, and $a(\cdot)$, $b(\cdot)$ and $c(\cdot)$ are functions
specific to the family. The conditional mean of $y_{ij}$, $\mu
_{ij}=E(y_{ij}|u_i)$, is assumed to depend on the fixed and random
effects through the linear predictor,
\[
\eta_{ij}={X_{ij}^R}^T
\beta^R+{X_{ij}^G}^T
\beta^G + {X_{ij}^R}^T
u_i
\]
with $g(\mu_{ij})=\eta_{ij}$ for some known link function, $g(\cdot)$.
Here, $X_{ij}^R$ and $X_{ij}=[{X_{ij}^R}^T, {X_{ij}^G}^T]^T$ are $r
\times1$ and $p \times1$ vectors of covariates and $\beta=[{\beta
^R}^T, {\beta^G}^T]^T$ is a $p \times1$ vector of fixed effects. We
considered the above breakdown (see \cite{Zhaetal06}) for the
linear predictor to allow for centering. For the $i$th cluster, let
$y_i=[y_{i1},\ldots, y_{in_i}]^T$,
$X_i^R=[X_{i1}^R,\ldots,X_{in_i}^R]^T$,
$X_i^G=[X_{i1}^G,\ldots,X_{in_i}^G]^T$,
$X_i=[X_{i1},\ldots,X_{in_i}]^T$ and $\eta_i=[\eta_{i1},\break\ldots,
\eta_{in_i}]^T$. Let $1_{n_i}$ denote the ${n_i} \times1$ column vector
with all entries equal to 1. We assume that the first column of $X_i^R$
is $1_{n_i}$ if $X_i^R$ is not a zero matrix. For Bayesian inference,
we specify prior distributions on the fixed effects $\beta$ and random
effects covariance matrix $D$. In this paper, we focus on responses
from the Bernoulli and Poisson families and the dispersion parameter is
one in these cases, so we do not consider a prior for $\phi$. We assume
a diffuse prior, $N(0,\Sigma_\beta)$, for $\beta$ and an independent
inverse Wishart prior, $\mathit{IW}(\nu,S)$, for~$D$. Following the suggestion
by \citet{KasNat06}, we set $\nu=r$ and let the scale matrix
$S$ be determined from first-stage data variability. In particular,
$S=r\hat{R}$, where
%
\begin{equation}\label{IWprior}
\hat{R}=c \Biggl(\frac{1}{n}\sum_{i=1}^{n}{X_i^R}^T
M_i(\hat{\beta })X_i^R
\Biggr)^{-1},
\end{equation}
$M_i(\hat{\beta})$ denotes the $n_i \times n_i$ diagonal generalized
linear model weight\vspace*{1pt} matrix with diagonal elements $[\phi v(\hat{\mu
}_{ij})\cdot g'(\hat{\mu}_{ij})^2]^{-1}$, $v(\cdot)$ is the variance
function based on $f(\cdot)$ in (\ref{GLMM}) and $g(\cdot)$ is the link
function. Here, $\hat{\mu}_{ij}=g^{-1}(X_{ij}^T\hat{\beta
}+{X_{ij}^R}^T\hat{u}_i)$, where $\hat{u}_i$ is set as $0$ for all $i$
and $\hat{\beta}$ is an estimate of the regression coefficients from
the generalized linear model obtained by pooling all data and setting
$u_i=0$ for all $i$. The value of $c$ is an inflation factor
representing the amount by which within-cluster variability should be
increased in determining $\hat{R}$. We used $c=1$ in all examples.

\section{A Partially Noncentered Parametrization for the Generalized
Linear Mixed Model}\label{sec3}
We introduce the following partially noncentered parametrization for
the GLMM. For each $i=1,\ldots,n$, the linear predictor is $\eta_i= X_i^R
\beta^R + X_i^G\beta^G + X_i^R u_i$. Let
\begin{eqnarray*}
X_i^G\beta^G &=& X_i^{G_1}
\beta^{G_1}+X_i^{G_2}\beta^{G_2}
\\
&=& 1_{n_i} {x_i^{G_1}}^T
\beta^{G_1} + X_i^{G_2}\beta^{G_2},
\end{eqnarray*}
where $\beta^{G_1}$ is a vector of length $g_1$ consisting of all
parameters corresponding to subject specific covariates (i.e., the
rows of $X_i^{G_1}$ are all the same and equal to the vector
$x_i^{G_1}$ say). Recall\vspace*{1pt} that the first column of $X_i^R$ is $1_{n_i}$
if $X_i^R$ is not a zero matrix. We have
\[
\eta_i = X_i^R \bigl( C_i
\beta^{RG_1}+ u_i\bigr) + X_i^{G_2}
\beta^{G_2},
\]
where
\[
C_i =\left[\matrix{&
{x_i^{G_1}}^T
\cr
I_r &
\cr
& 0 }
\right] \quad\mbox{and}\quad \beta^{RG_1}= \left[\matrix{\beta^R
\cr
\beta^{G_1}}\right].
\]

Let $\alpha_i = C_i \beta^{RG_1}+ u_i$ and $\tilde{\alpha}_i=\alpha
_i-W_i C_i \beta^{RG_1}$,
where $W_i$ is an $r \times r$ matrix to be specified. The proportion
of $C_i \beta^{RG_1}$ subtracted from each $\alpha_i$ is allowed to
vary with $i$ as in Papaspiliopoulos, Roberts and Sk{\"o}ld (\citeyear{PapRobSko03}) to reflect the
varying informativity of each response $y_i$ about the underlying
$\alpha_i$. $W_i=0$ corresponds to the centered and $W_i=I$ to the
noncentered parametrization. Finally,
%
\begin{eqnarray}\label{partiallynoncenteredparametrization}
\eta_i &=& X_i^R \bigl(\tilde{
\alpha}_i+W_i C_i \beta^{RG_1}
\bigr)+ X_i^{G_2}\beta ^{G_2}
\nonumber\\[-8pt]\\[-8pt]
&=& V_i \beta+X_i^R \tilde{
\alpha}_i,\nonumber
\end{eqnarray}
where $V_i = [ X_i^R W_iC_i\enskip X_i^{G_2}]$ and $\tilde{\alpha}_i \sim
N ((I-W_i)\cdot C_i \beta^{RG_1},D )$. We refer to (\ref{partiallynoncenteredparametrization}) as the partially noncentered
parametrization. Let $\tilde{\alpha}=[\tilde{\alpha}_1^T,\ldots,\tilde
{\alpha}_n^T]^T$ and $\theta=\{\beta,D,\tilde{\alpha}\}$ denote the set\vadjust{\goodbreak}
of unknown parameters in the GLMM.
The joint distribution of $p(y,\theta)$ is
%
\begin{eqnarray}\label{factorization}
p(y,\theta) &=& \Biggl\{ \prod_{i=1}^n
p(y_i|\beta,\tilde{\alpha}_i)p(\tilde {
\alpha}_i|\beta,D) \Biggr\}\nonumber\\[-8pt]\\[-8pt]
&&{}\cdot  p(\beta|\Sigma_\beta) p(D|
\nu,S).\nonumber
\end{eqnarray}
Figure~\ref{FG} shows the factor graph for $p(y,\theta)$ where there is
a node (circle) for every variable, which is shaded in the case of
observed variables, and a node (filled rectangle) for each factor in
the joint distribution. Constants or hyperparameters are denoted with
smaller filled circles. Each factor node is connected by undirected
links to all of the variable nodes on which that factor depends (see
\cite{Bis06}).
\setcounter{figure}{0}
\begin{figure}

\includegraphics{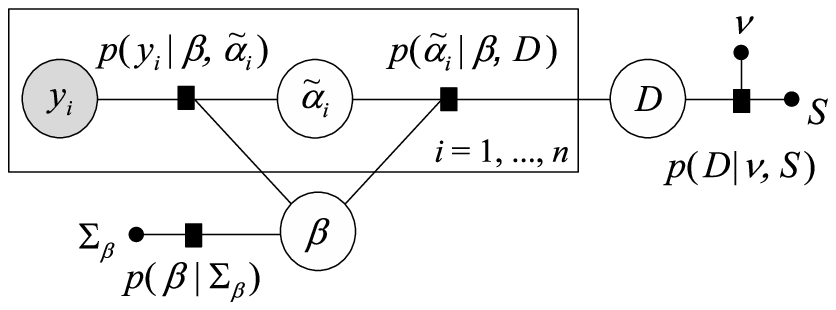}

\caption{Factor graph for $p(y,\theta)$ in (\protect\ref
{factorization}). Filled rectangles denote factors and circles denote
variables (shaded for observed variables). Smaller filled circles
denote constants or hyperparameters. The box represents a plate which
contains variables and factors to be replicated. Number of repetitions
is indicated in the lower right corner.}
\label{FG}
\end{figure}
Next, we consider specification of the tuning parameter $W_i$,
referring to the linear mixed mod\-el example in Section~\ref{sec1.1} which is a
special case of the GLMM in (\ref{GLMM}) with an identity link.

\subsection{Specification of Tuning Parameter}\label{sec3.1}
It is interesting to note that for the linear mixed model in (\ref
{linearmixedmodel}), the expression for $W_i$ leading to optimal
performance in VB and the Gibbs sampling algorithm is exactly the same
(see Papaspiliopoulos, Roberts and Sk{\"o}ld, \citeyear{PapRobSko03}).  Gelfand, Sahu and Carlin (\citeyear{GelSahCar95})
also observed the importance of $W_i$ in assessing convergence
properties of the centered parametrization. They showed that $|W_i|<1$
for all $i$ and $|W_i|$ is close to zero (centering is more efficient)
when the generalized variance $|D|$ is large. On the other hand,
$|W_i|$ is close to 1 (noncentering works better) when the error
variance is large. Outside the Gaussian context,
Papaspiliopoulos, Roberts and\break Sk{\"o}ld (\citeyear{PapRobSko03}) considered partial noncentering for the spatial GLMM and
specified the tuning parameters by using a quadratic expansion of the
log-likelihood to obtain an indication of the information\vadjust{\goodbreak} present in~$y_i$. Observe that $W_i$ in (\ref{Wi}) can be expressed as
%
\begin{equation}\label{Wiexp}
W_i= \bigl(\mathcal{I}_f +D^{-1}
\bigr)^{-1}D^{-1},
\end{equation}
if $\ell=\log p(y_i|\beta,\alpha_i)$ denotes the log-likelihood and
$\mathcal{I}_f=-\frac{\partial^2 \ell}{\partial{\alpha}_i\,\partial{\alpha}_i^T}$.
We use (\ref{Wiexp}) to extend partially noncentered parametrizations
to GLMMs and consider the specification of $W_i$ for responses from the
Bernoulli and Poisson families in particular.

Recall that the linear predictor $\eta_i$ can be expressed as
$X_i^R\alpha_i+X_i^{G_2}\beta^{G_2}$. For Poisson responses with the
log link function, we allow for an offset $\log E_{ij}$ so that $\log
\mu_{ij}=\log E_{ij}+ \eta_{ij}$. Let $E_i=[E_{i1},\ldots,E_{in_i}]^T$.
We have
%
\begin{eqnarray}\label{Poisson}\qquad
\ell&=& y_i^T (\log E_i+\eta_i
)-E_i^T\exp(\eta_i)\nonumber\\
&&{}-1_{n_i}^T
\log (y_i!)\quad\mbox{and}\\
\mathcal{I}_f&=& \sum_{j=1}^{n_i}E_{ij}
\exp(\eta _{ij}){X_{ij}^R} {X_{ij}^R}^T
\approx\sum_{j=1}^{n_i}y_{ij}{X_{ij}^R}
{X_{ij}^R}^T,
\nonumber
\end{eqnarray}
if we approximate the conditional mean $\mu_{ij}$ with the response.
For Bernoulli responses with the logit link function, we have
%
\begin{eqnarray}\label{logistic}
\ell&=& y_i^T \eta_i-1_{n_i}^T
\log \bigl\{1_{n_i}+\exp(\eta_i) \bigr\}
\quad\mbox{and}\nonumber\\[-8pt]\\[-8pt]
\mathcal{I}_f&=& \sum_{j=1}^{n_i}
\frac{\exp(\eta
_{ij})}{\{1+
\exp(\eta_{ij}) \}^2}{X_{ij}^R} {X_{ij}^R}^T.\nonumber
\end{eqnarray}
The specification of $W_i$ depends on the random effects covariance $D$
and, for Bernoulli responses, on the linear predictor $\eta_i$ as well.
In Algorithm~\ref{algor3}, we initialize $W_i$ by considering $\eta_i=X_i\beta
+X_i^Ru_i$ and using estimates of $D$, $\beta$ and $u_i$ from penalized
quasi-likelihood. Subsequently, we can either keep $W_i$ as fixed or
update them by replacing $D$ with $\frac{S^q}{\nu^q-r-1}$, assuming the
variational posterior\vspace*{1pt} of $D$ is $\mathit{IW}(\nu^q,S^q)$ and $\eta_i$ with $V_i
\mu_\beta^q+X_i^R \mu_{\tilde{\alpha}_i}^q$, where $\mu_\beta^q$ and
$\mu_{\tilde{\alpha}_i}^q$ are the variational posterior means of $\beta
$ and $\tilde{\alpha}_i$, respectively. This can be done at the
beginning of each iteration after new estimates of $\mu_\beta^q$, $\mu
_{\tilde{\alpha}_i}^q$, $\nu^q$ and $S^q$ are obtained (see Algorithm
\ref{algor3}
step 1).

\section{Variational Inference for GLMMs}\label{sec4}
In this section we present the nonconjugate variational message passing
algorithm recently developed in machine learning by \citet{KnoMin}
for fitting GLMMs. Recall that in VB, the posterior distribution
$p(\theta|y)$ is approximated by a $q(\theta)$ which is assumed to be
of a factorized form, say, $q(\theta)=\prod_{i=1}^m q_i(\theta_i)$ for
some partition $\{\theta_1,\ldots,\theta_m\}$ of~$\theta$. For
conjugate-exponential\vadjust{\goodbreak} models, the optimal densities $q_i$ will have the
same form as the prior so that it suffices to update the parameters of
$q_i$, such as in Algorithm~\ref{algor1}. Variational message passing
(\cite{WinBis05}) is an algorithm which allows VB to be applied to
conjugate-exponential models without having to derive
application-specific updates. In the case of GLMMs where the responses
are from the Bernoulli or Poisson families, the factor
$p(y_i|\beta,\tilde{\alpha}_i)$ of $p(y,\theta)$ in (\ref
{factorization}) is nonconjugate with respect to the prior
distributions over $\beta$ and $\tilde{\alpha}_i$ for each
$i=1,\ldots,n$. Therefore, if we apply VB and assume, say,
$q(\theta)=q(\beta)q(D)\prod_{i=1}^n q(\tilde{\alpha}_i)$, the optimal
densities for $q(\beta)$ and $q(\tilde{\alpha}_i)$ will not belong to
recognizable density families.

\subsection{Nonconjugate Variational Message Passing}\label{sec4.1}
In nonconjugate variational message passing, besides assuming that
$q(\theta)$ must factorize into\break $\prod_{i=1}^m q_i(\theta_i)$ for some
partition $\{\theta_1,\ldots,\theta_m\}$ of $\theta$, we impose another
restriction that each $q_i$ must belong to some exponential family. In
this way, we only have to find the parameters of each $q_i$ that
maximizes the lower bound $\mathcal{L}$. Suppose each $q_i$ can be
written in the form
\[
q_i(\theta_i)=\exp\bigl\{\lambda_i^T
t(\theta_i)-h(\lambda_i)\bigr\},
\]
where $\lambda_i$ is the vector of natural parameters and $t(\cdot)$
are the sufficient statistics. We wish to maximize $\mathcal{L}$ with
respect to the variational parameters $\lambda_1,\ldots,\lambda_m$ which
are also natural parameters of\break $q_1(\theta_1),\ldots,q_m(\theta_m)$,
respectively. In the following, we show that nonconjugate variational
message passing can be interpreted as a fixed-point iteration where
updates are obtained from the condition that the gradient of $\mathcal
{L}$ with respect to each $\lambda_i$ is zero when $\mathcal{L}$ is maximized.

From (\ref{lowerbound}), the gradient of $\mathcal{L}$ with respect to
$\lambda_i$ is
%
\begin{equation}\label{usualgradient}
\frac{\partial\mathcal{L}}{\partial\lambda_i} = \frac{\partial
}{\partial\lambda_i}E_q\bigl\{\log p(y,\theta)
\bigr\}-\frac{\partial}{\partial
\lambda_i}E_q\bigl\{\log q(\theta)\bigr\}.\hspace*{-28pt}
\end{equation}
Let us consider the first term in (\ref{usualgradient}). Suppose
$p(y,\break\theta)=\prod_a f_a(y,\theta)$. We have
\[
E_q\bigl\{\log p(y,\theta)\bigr\}=\sum_a
S_a,
\]
where
\[
S_a = E_q\bigl\{ \log
f_a(y,\theta)\bigr\}.
\]
Note that each $S_a$ is a function of the natural parameters $\lambda
_1,\ldots,\lambda_m$. Since we have assumed that $\theta_i$ is independent
of all $\theta_j$ where $j \neq i$ in the variational approximation
$q$, the only terms in $\sum_a S_a$ which depend on $\lambda_i$ are the
factors $f_a$ connected\vadjust{\goodbreak} to $\theta_i$ in the factor graph of $p(y,\theta
)$. Therefore,
%
\begin{equation}\label{firstterm}
\frac{\partial}{\partial\lambda_i}E_q\bigl\{\log p(y,\theta)\bigr\}= \sum
_{a
\in N(\theta_i)} \frac{\partial S_a}{\partial\lambda_i},
\end{equation}
where the summation is over all factors in $N(\theta_i)$, the
neighborhood of $\theta_i$ in the factor graph. For the second term in
(\ref{usualgradient}), we have
\[
E_q\bigl\{\log q(\theta)\bigr\}= \sum_{l=1}^m
E_q\bigl\{\log q_l(\theta_l)\bigr\},
\]
where the only term in the sum that depends
on $\lambda_i$ is the $i$th term. Hence,
%
\begin{eqnarray}\label{secondterm}\qquad
\frac{\partial}{\partial\lambda_i} E_q\bigl\{\log q(\theta)\bigr\} &=&
\frac{\partial}{\partial\lambda_i} \biggl\{\lambda_i^T\,
\frac
{\partial h(\lambda_i)}{\partial\lambda_i}-h(\lambda_i) \biggr\}
\nonumber\\[-8pt]\\[-8pt]
&=& \mathcal{V}(\lambda_i) \lambda_i,\nonumber
\end{eqnarray}
where we have used the fact that $E_q\{t(\theta_i)\}=\frac{\partial
h(\lambda_i)}{\partial\lambda_i}$ and $\mathcal{V}(\lambda_i)=\frac
{\partial^2 h(\lambda_i)}{\partial\lambda_i \,\partial\lambda_i^T}$
denotes the variance--covariance matrix of $t(\theta_i)$. Note that
$\mathcal{V}(\lambda_i)$ is symmetric positive semi-definite. Putting
(\ref{firstterm}) and (\ref{secondterm}) together, the gradient of
the lower bound is
\[
\frac{\partial\mathcal{L}}{\partial\lambda_i}=\sum_{a \in N(\theta
_i)}\frac{\partial S_a}{\partial\lambda_i}-
\mathcal{V}(\lambda_i) \lambda_i
\]
and is zero when $\lambda_i = \mathcal{V}(\lambda_i)^{-1}\sum_{a \in
N(\theta_i)} \frac{\partial S_a}{\partial\lambda_i}$, provided
$\mathcal{V}(\lambda_i)$ is invertible. This condition is used to
obtain updates to $\lambda_i$ in nonconjugate variational message
passing (Algorithm~\ref{algor2}).

{\setcounter{figure}{1}
\renewcommand{\figurename}{Algorithm}
%
\begin{figure}
\begin{tabular}{@{}l@{}}
\rule{\columnwidth}{0.5pt}\\
Initialize $\lambda_i$ for $i=1,\ldots,m$.\\
Cycle:\\
\hphantom{Cycle:}For $i=1,\ldots,m$,\\
\hphantom{Cycle:}$\lambda_i \leftarrow \mathcal{V} (\lambda_i )^{-1}\sum_{a \in N(\theta_i)} \frac{\partial S_a}{\partial\lambda_i}$ \\
until convergence.
\\[-6pt]
\rule{\columnwidth}{0.5pt}
\end{tabular}
\caption{Nonconjugate variational message passing.}\label{algor2}
\end{figure}
}

The updates can be simplified when the factor $f_a$ is conjugate to
$q_i(\theta_i)$, that is, $f_a$ has the same functional form as
$q_i(\theta_i)$ with respect to $\theta_i$. Let $\theta_{-i}=(\theta
_1,\ldots,\theta_{i-1},\theta_{i+1},\ldots,\theta_m)$. Suppose
\[
f_a(y,\theta)= \exp\bigl\{g_a(y,\theta_{-i})^T
t(\theta_i)-h_a(y,\theta _{-i})\bigr\}.
\]
Then $\frac{\partial S_a}{\partial\lambda_i} = \mathcal{V}(\lambda
_i)E_q\{g_a(y,\theta_{-i})\}$,
where $E_q\{g_a(y,\break\theta_{-i})\}$ does not depend on $\lambda_i$. When
every factor in the neighborhood of $\theta_i$ is conjugate to
$q_i(\theta_i)$, the gradient of the lower bound can be simplified to\break
$\mathcal{V}(\lambda_i)  [ \sum_{a \in N(\theta_i)}E_q\{g_a(y,\theta
_{-i})\}- \lambda_i  ]$
and the updates in nonconjugate variational message passing reduce\vadjust{\goodbreak} to
%
\begin{equation}\label{VMP}
\lambda_i \leftarrow\sum_{a \in N(\theta_i)}E_q
\bigl\{g_a(y,\theta_{-i})\bigr\}.
\end{equation}
These are precisely the updates in variational message passing.
Nonconjugate variational message passing thus reduces to variational
message passing for conjugate factors (see also Knowles and Minka\break (\citeyear{KnoMin})). Unlike
variational message passing, however, the Kullback--Leibler divergence
is not guaranteed to decrease at each step and sometimes convergence
problems may be encountered. \citet{KnoMin} suggested using
damping to fix convergence problems. We did not encounter any
convergence problems in the examples considered in this paper.

\subsection{Updates for Multivariate Gaussian Variational Distribution}\label{sec4.2}
Suppose $q_i$ is Gaussian. While the updates in Algorithm~\ref{algor2} are
expressed in terms of the natural parameters~$\lambda_i$, it might be
more convenient to express $\frac{\partial S_a}{\partial\lambda_i}$
in terms of the mean and covariance of~$q_i$. \citet{KnoMin}
have considered the univariate case and \citet{Wan} derived fully
simplified updates for the multivariate case. However, as \citet{Wan}
is in preparation, we give enough details of the update so that the
derivation can be understood. \citet{MagNeu88} is a good
reference for the matrix differential calculus techniques used in the
derivation.

Suppose $q_i(\theta_i)=N(\mu_{\theta_i}^q,\Sigma_{\theta_i}^q)$ where
$\theta_i$ is a vector of length $d$. For a $d \times d$ square matrix
$A$, $\operatorname{vec}(A)$ denotes the $d^2 \times1$ vector obtained
by stacking the columns of $A$ under each other, from left to right in
order, and $\operatorname{vech}(A)$ denotes the $\frac{1}{2}d(d+1)
\times1$
vector obtained from $\operatorname{vec}(A)$ by eliminating all supradiagonal
elements of $A$. We can write $q_i(\theta_i)$ as
\[
\exp\lleft\{ \lambda_i^T \left[\matrix{
\operatorname{vech}\bigl(\theta_i\theta_i^T
\bigr)
\cr
\theta_i}\right] -h(\lambda_i) \rright\}
\]
where
\[
\lambda_i=\left[\matrix{-\frac{1}{2}D_d^T
\operatorname{vec}\bigl({\Sigma_{\theta_i}^q}^{-1}
\bigr)
\cr
{\Sigma_{\theta_i}^q}^{-1}
\mu_{\theta_i}^q}\right]
\]
and $h(\lambda_i)= \frac{1}{2}
{\mu_{\theta_i}^q}^T{\Sigma_{\theta_i}^q}^{-1}\mu_{\theta_i}^q
+{\frac{1}{2}\log}|\Sigma_{\theta_i}^q|+\frac{d}{2}\log(2\pi)$. The
matrix $D_d$ is a unique $d^2 \times\frac{1}{2}d(d+1)$ matrix that
transforms $\operatorname{vech}(A)$ into $\operatorname{vec}(A)$ if $A$
is symmetric,
that is, $D_d\operatorname{vech}(A)=\operatorname{vec}(A)$. Let $D_d^+$
denote the
Moore--Penrose inverse of $D_d$. If we let
$\lambda_{i1}=-\frac{1}{2}D_d^T \operatorname{vec}({\Sigma_{\theta_i}^q}^{-1})$
and\vadjust{\goodbreak} $\lambda_{i2} = {\Sigma_{\theta_i}^q}^{-1}\mu_{\theta_i}^q$,
$\frac{\partial S_a}{\partial\lambda_i}$ can be expressed as
\begin{eqnarray*}
\left[\matrix{ \displaystyle \frac{\partial S_a}{\partial\lambda_{i1}}
\vspace*{2pt}\cr \displaystyle \frac{\partial S_a}{\partial\lambda_{i2}} }\right] &=&
\left[ \matrix{ \displaystyle
\frac{\partial\operatorname{vec}(\Sigma_{\theta_i}^q)}{\partial\lambda
_{i1}} & \displaystyle
\frac{\partial\mu_{\theta_i}^q}{\partial\lambda_{i1}} \vspace*{2pt}\cr \displaystyle
\frac{\partial\operatorname{vec}(\Sigma_{\theta_i}^q)}{\partial\lambda
_{i2}} & \displaystyle
\frac{\partial\mu_{\theta_i}^q}{\partial\lambda_{i2}}}\right]
\left[\matrix{ \displaystyle \frac{\partial
S_a}{\partial\operatorname{vec}(\Sigma_{\theta_i}^q)} \vspace*{2pt}\cr
\displaystyle \frac{\partial S_a}{\partial\mu_{\theta_i}^q}}\right] \\
&=& U(\lambda_i)
\left[\matrix{ \displaystyle \frac{\partial
S_a}{\partial\operatorname{vec}(\Sigma_{\theta_i}^q)} \vspace*{2pt}\cr
\displaystyle \frac{\partial S_a}{\partial\mu_{\theta_i}^q} }\right],
\end{eqnarray*}
where
\[
U(\lambda_i) = \left[\matrix{ 2D_d^+
\bigl(\Sigma_{\theta_i}^q\otimes\Sigma_{\theta_i}^q
\bigr) & 2D_d^+\bigl(\mu_{\theta_i}^q\otimes
\Sigma_{\theta_i}^q\bigr)
\cr
0 &\Sigma_{\theta_i}^q}
\right]
\]
and $\otimes$ denotes the Kronecker product. Moreover, $\mathcal
{V}(\lambda_i)=\frac{\partial^2 h(\lambda_i)}{\partial\lambda_i\,
\partial\lambda_i^T}$ can be derived to be
%
\[
\left[
\begin{array}{l@{\quad}c}
2D_d^+\bigl(\mu_{\theta_i}^q
{\mu_{\theta_i}^q}^T \otimes\Sigma_{\theta
_i}^q\\
\hspace*{25pt}{}+\Sigma_{\theta_i}^q \otimes \mu_{\theta_i}^q {
\mu_{\theta_i}^q}^T& 2D_d^+\bigl(\mu_{\theta_i}^q\otimes
\Sigma_{\theta_i}^q\bigr)\\
\hspace*{47pt}{}+\Sigma_{\theta_i}^q
\otimes\Sigma _{\theta_i}^q\bigr){D_d^+}^T
\\
\hspace*{20pt}\bigl\{2D_d^+\bigl(
\mu_{\theta_i}^q\otimes\Sigma_{\theta_i}^q\bigr)
\bigr\}^T & \Sigma_{\theta_i}^q
\end{array}\right].
\]
The update for $\lambda_i$ can be computed as
\[
\lambda_i \leftarrow\mathcal{V}(\lambda_i)^{-1}
U(\lambda_i) \sum_{a
\in N(\theta_i)} \left[\matrix{
\displaystyle \frac{\partial S_a}{\partial\operatorname{vec}(\Sigma_{\theta_i}^q)}
\vspace*{2pt}\cr
\displaystyle \frac{\partial S_a}{\partial\mu_{\theta_i}^q} }\right]
\]
and
\[
\mathcal{V}(
\lambda_i)^{-1} U(\lambda_i)= \left[\matrix{
D_d^T & 0
\cr
-2\bigl({\mu_{\theta_i}^q}^T \otimes I
\bigr){D_d^+}^T D_d^T & I }
\right].
\]
\citet{Wan} showed that the updates simplify to
%
\begin{eqnarray}\label{Algorithm3updates}
\Sigma_{\theta_i}^q &\leftarrow&-\frac{1}{2} \biggl[
\operatorname {vec}^{-1} \biggl(\sum_{a \in N(\theta_i)}
\frac{\partial
S_a}{\partial\operatorname{vec}(\Sigma_{\theta_i}^q)} \biggr) \biggr]^{-1}
\quad\mbox{and}\hspace*{-22pt}\nonumber\\[-8pt]\\[-8pt]
\mu_{\theta_i}^q
&\leftarrow&\mu_{\theta_i}^q + \Sigma _{\theta_i}^q
\sum_{a \in N(\theta_i)} \frac{\partial
S_a}{\partial\mu_{\theta_i}^q}.\hspace*{-22pt}\nonumber
\end{eqnarray}
A more detailed version of the argument will be given in the
forthcoming manuscript of \citet{Wan}.

{\setcounter{figure}{2}
\renewcommand{\figurename}{Algorithm}
%
\begin{figure*}
\begin{tabular}{@{}l@{}}
\rule{394pt}{0.5pt}\\
Initialize $\mu_{\beta}^q$, $\Sigma_{\beta}^q$, $S^q$ and $\mu_{\tilde
{\alpha}_i}^q$, $\Sigma_{\tilde{\alpha}_i}^q$, $W_i$ for $i=1,\ldots,n$
and set $\nu^q = n+\nu$.\\
Cycle: \\
\hphantom{C}1. Update $W_i$ and hence $V_i$ for $i=1,\ldots,n$. (Optional)
\\
\hphantom{C}2. $\Sigma_\beta^q \leftarrow (\Sigma_\beta^{-1} + \nu
^q\sum_{i=1}^n {\tilde{W}_i}^T{S^q}^{-1}\tilde{W}_i +\sum_{i=1}^n \sum_{j=1}^{n_i} F_{ij}V_{ij}V_{ij}^T  )^{-1}$ \\
\hphantom{C2. }$\mu_\beta^q \leftarrow\mu_\beta^q + \Sigma_\beta^q
\{-\Sigma_\beta^{-1}\mu_\beta^q + \nu^q\sum_{i=1}^n {\tilde
{W}_i}^T{S^q}^{-1}(\mu_{\tilde{\alpha}_i}^q-\tilde{W}_i\mu_\beta^q) +
\sum_{i=1}^n V_i^T(y_i-G_i)  \}$ \\
\hphantom{C}3. For $i=1,\ldots,n$,\\
\hphantom{C2. }$\Sigma_{\tilde{\alpha}_i}^q \leftarrow (\nu^q
{S^q}^{-1}+\sum_{j=1}^{n_i} F_{ij} X_{ij}^R {X_{ij}^R}^T
)^{-1}$ \\
\hphantom{C2. }$\mu_{\tilde{\alpha}_i}^q \leftarrow\mu_{\tilde{\alpha
}_i}^q + \Sigma_{\tilde{\alpha}_i}^q  \{-\nu^q{S^q}^{-1}(\mu_{\tilde
{\alpha}_i}^q-\tilde{W}_i\mu_\beta^q) + {X_i^R}^T (y_i-G_i)  \} $ \\
\hphantom{C}4. $S^q \leftarrow S+\sum_{i=1}^n  \{ (\mu_{\tilde
{\alpha}_i}^q-\tilde{W}_i \mu_\beta^q) (\mu_{\tilde{\alpha}_i}^q-\tilde
{W}_i \mu_\beta^q)^T + \Sigma_{\tilde{\alpha}_i}^q+\tilde{W}_i\Sigma
_\beta^q\tilde{W}_i^T  \}$ \\
until the absolute relative change in the lower bound $\mathcal{L}$ is
negligible.
\\[-6pt]
\rule{394pt}{0.5pt}
\end{tabular}
\caption{Nonconjugate variational message passing for fitting GLMMs.}
\label{algor3}
\end{figure*}
}

\subsection{Nonconjugate Variational Message Passing Algorithm for
Generalized Linear Mixed Models}\label{sec4.3}
For the GLMM, we consider a variational approximation of the form
%
\begin{equation}\label{VAGLMM}
q(\theta)=q(\beta)q(D)\prod_{i=1}^n q(
\tilde{\alpha}_i),
\end{equation}
where $q(\beta)$ is $N (\mu_{\beta}^q,\Sigma_{\beta}^q )$,
$q(D)$ is $\mathit{IW}(\nu^q,S^q)$, and $q(\tilde{\alpha}_i)$ is $N (\mu
_{\tilde{\alpha}_i}^q,\Sigma_{\tilde{\alpha}_i}^q )$, all
belonging to the exponential family. Here, we approximate the posterior
distributions of $\beta$ and $\tilde{\alpha}_i$ by Gaussian
distributions which are often reasonable and supported by the
asymptotic normality of the posterior. Our results also indicate that
Gaussian approximation performs reasonably well as an approximation to
the posterior in finite samples. See \citet{Geletal04} for
further discussion and counterexamples. The posterior distribution for
$D$ is approximated by an inverse Wishart which can be shown to be the
optimal density under only the VB assumption $q(\theta)=q(\beta)q(D)
q(\tilde{\alpha})$. The nonconjugate variational message passing
algorithm for GLMMs is outlined in Algorithm~\ref{algor3}.

In Algorithm~\ref{algor3}, for each $i=1,\ldots,n$, $j=1,\ldots,n_i$,
$\tilde{W}_i=[(I-W_i)C_i \enskip0_{r\times(p-r-g_1)}]$,
$\kappa_{ij}$ is the $j$th component\vspace*{-1pt} of $\kappa_i=\exp
\{V_i \mu_\beta^q +X_i^R
\mu_{\tilde{\alpha}_i}^q+\frac{1}{2}\operatorname{diag}(V_i
\Sigma_\beta^q {V_i}^T+X_i^R\Sigma_{\tilde{\alpha}_i}^q{X_i^R}^T)\}$,
$\mu_{ij}$ is the $j$th component of $\mu_i=V_{i}\mu_\beta^q +X_{i}^R
\mu_{\tilde{\alpha}_i}^q$, $\sigma_{ij}$ is the $j$th component of
$\sigma_i=\sqrt{\operatorname{diag}(V_{i} \Sigma_\beta^q V_{i}^T
+X_{i}^R \Sigma_{\tilde{\alpha}_i}^q {X_{i}^R}^T)}$ and
$B^{(r)}(\mu,\sigma)=\break\int_{-\infty}^{\infty} b^{(r)}(\sigma
x+\mu)\frac{1}{\sqrt{2\pi}}\mathrm{e}^{-x^2} \,\mathrm{d}x$ where
$b(x)=\log(1+\mathrm{e}^x)$ and $b^{(r)}(x)$ denotes the $r$th
derivative of $b(\cdot)$ with respect to $x$. If $\mu$ and $\sigma$ are
vectors, say,
\[
\mu=\left[\matrix{1
\cr
2
\cr
3} \right] \quad\mbox{and}\quad \sigma=\left[\matrix{4
\cr
5
\cr
6} \right],
\]
then
\[
B^{(r)} (\mu,\sigma)=\left[\matrix{B^{(r)}(1,4)
\vspace*{2pt}\cr
B^{(r)}(2,5)
\vspace*{2pt}\cr
B^{(r)}(3,6)} \right].
\]
In addition,
\[
F_{ij}= \cases{E_{ij} \kappa_{ij}, & if Poisson,
\vspace*{2pt}\cr
B^{(2)}(\mu_{ij}, \sigma_{ij}), & if logistic,}
\]
and
\[
G_i = \cases{ E_i\odot\kappa_i,
& if Poisson,
\vspace*{2pt}\cr
B^{(1)}(\mu_i, \sigma_i), &
if logistic,}
\]
where $a\odot b$ denotes the element-wise product of two vectors, $a$
and $b$.

The updates in Algorithm~\ref{algor3} can be obtained from the formulae in (\ref
{VMP}) and (\ref{Algorithm3updates}). Consider the parameters $\nu^q$
and $S^q$ of $q(D)$. The factors connected to $D$ are $p(D|\nu,S)$ and
$p(\tilde{\alpha}_i|\beta,D)$, $i=1,\ldots,n$, which are all conjugate
factors. Therefore, updates for $q(D)$ can be obtained from (\ref{VMP})
or by setting $q(D) \propto\exp\{E_{-D} \log p(y,\theta)\}$ as in VB.
The shape parameter $\nu^q$ can be shown to be deterministic: $\nu^q =
n+\nu$ and the update for $S^q$ is given in step 4 of Algorithm~\ref{algor3}. The
updates of the parameters of $q(\beta)$ and $q(\tilde{\alpha}_i)$,
$i=1,\ldots,n$, have to be computed using (\ref{Algorithm3updates}),
as $p(y_i|\beta,\tilde{\alpha}_i)$ connected to $\beta$ and
$\tilde{\alpha}_i$ is a nonconjugate factor. The factors connected to
$\beta$ are $p(\beta|\Sigma_\beta)$, $p(\tilde{\alpha}_i|\beta,D)$ and
$p(y_i|\beta,\tilde{\alpha}_i)$, $i=1,\ldots,n$ (see Figure~\ref{FG}).
Let $S_{\beta}=E_q \{\log p(\beta|\Sigma_\beta)\}$,
$S_{\tilde{\alpha}_i}= E_q \{ \log p(\tilde{\alpha}_i|\beta,D) \}$ and
$S_{y_i}= E_q\{\log p(y_i|\beta,\tilde{\alpha}_i)\}$, $i=1,\ldots,n$,
where $E_q$ denotes expectation with respect to $q$. We have
\begin{eqnarray*}
\sum_{a \in N(\beta)} \frac{\partial S_a}{\partial\operatorname
{vec}(\Sigma_{\beta}^q)} &=&\frac{\partial S_{\beta}}{\partial
\operatorname{vec}(\Sigma_{\beta}^q)}+
\sum_{i=1}^n \frac{\partial
S_{\tilde{\alpha}_i}}{\partial\operatorname{vec}(\Sigma_{\beta}^q)} \\
&&{}+ \sum
_{i=1}^n \frac{\partial S_{y_i}}{\partial
\operatorname{vec}(\Sigma_{\beta}^q)},
\\
\sum_{a \in N(\beta)} \frac{\partial S_a}{\partial\mu_{\beta}^q}
&=&\frac{\partial S_{\beta}}{\partial\mu_{\beta}^q} + \sum_{i=1}^n
\frac{\partial S_{\tilde{\alpha}_i}}{\partial\mu_{\beta}^q}+ \sum
_{i=1}^n \frac{\partial S_{y_i}}{\partial\mu_{\beta}^q},
\end{eqnarray*}
and the simplified updates for $\Sigma_\beta^q$ and $\mu_\beta^q$ are
given in step 2 of Algorithm~\ref{algor3}. The factors connected to $\tilde{\alpha
}_i$ are $p(\tilde{\alpha}_i|\beta,D)$ and $p(y_i|\beta,\tilde{\alpha
}_i)$ for $i=1,\ldots,n$ (see Figure~\ref{FG}). Hence,
\[
\sum_{a \in N(\tilde{\alpha}_i)}\frac{\partial S_a}{\partial
\operatorname{vec}(\Sigma_{\tilde{\alpha}_i}^q)} =\frac{\partial
S_{\tilde{\alpha}_i}}{\partial\operatorname{vec}(\Sigma_{\tilde{\alpha
}_i}^q)}+
\frac{\partial S_{y_i}}{\partial\operatorname{vec}(\Sigma
_{\tilde{\alpha}_i}^q)}
\]
and
\[
\sum_{a \in N(\tilde{\alpha}_i)}
\frac{\partial S_a}{\partial\mu
_{\tilde{\alpha}_i}^q} =\frac{\partial S_{\tilde{\alpha}_i}}{\partial
\mu_{\tilde{\alpha}_i}^q}+\frac{\partial S_{y_i}}{\partial\mu_{\tilde
{\alpha}_i}^q}.
\]
The simplified updates for $\Sigma_{\tilde{\alpha}_i}^q$ and $\mu
_{\tilde{\alpha}_i}^q$ are given in step 3 of Algorithm~\ref{algor3}. See Appendix
\ref{secA} for the evaluation of $S_\beta$, $S_{\tilde{\alpha}_i}$ and
$S_{y_i}$. All gradients can be computed using vector differential
calculus (see \cite{MagNeu88}).

For responses from the Poisson family, $S_{y_i}$ can be evaluated in
closed form. However, $S_{y_i}$ cannot be evaluated analytically for
Bernoulli responses. \citet{KnoMin} discussed several alternatives
in handling this integral. One could construct a bound on
$\log(1+\mathrm{e}^x)$ such as the ``quadratic'' bound (\cite{Jaa}) or
the ``tilted'' bound (\cite{Sau}). We observed a negative bias in the
estimates for the random effects variances when using the ``tilted
bound'' in Algorithm~\ref{algor3}. This negative bias decreases as the
cluster size increases (see also \cite{RijVom08}). Hence, we use
quadrature to compute the expectation and gradients. Following
\citet{OrmWan12}, we reduce all high-dimensional integrals to
univariate ones and evaluate these efficiently using adaptive
Gauss--Hermite quadrature (\cite{LiuPie94}). The details are given in
Appendix~\ref{secB}.

While the updates in Algorithm~\ref{algor1} can be simplified if $W_i=I$
(noncentered) or 0 (centered) and are more complex in the partially
noncentered case, the reduction in efficiency is minimal. Moreover,
with a good initialization, it is feasible to keep $W_i$ as fixed
throughout the course of running Algorithm~\ref{algor3} so that no additional
computation time is used in updating $W_i$. We use the fit from
penalized quasi-likelihood implemented via the function
\texttt{glmmPQL()} in the \texttt{R}
package \texttt{MASS} (\cite{VenRip02})
to initialize Algorithm~\ref{algor3}. In our experiments, the lower bound computed
at the end of each cycle of updates is usually on an increasing trend
although there might be some instability at the beginning. In cases
where the algorithm does not converge, we found that changing the
initialization can help to alleviate the situation. Although the lower
bound is not guaranteed to\vadjust{\goodbreak} increase at the end of each cycle, we
continue to use it as a means of monitoring convergence and Algorithm
\ref{algor3}
is terminated when the absolute relative change in the lower bound is
less than~$10^{-6}$. The lower bounds for the logistic and Poisson
GLMMs are presented in Appendix~\ref{secA}.

\section{Model Selection Based on Variational Lower Bound}\label{sec5}
At the point of convergence of Algorithm~\ref{algor3}, the lower bound on the log
marginal likelihood, $\log p(y)$, is maximized. This variational lower
bound is often tight and can be useful for model selection. Bayesian
model selection is traditionally based on computation of Bayes factor
in which marginal likelihood plays an important role. Suppose there are
$k$ candidate models, $M_1,\ldots, M_k$. Let $p(M_j)$ and $p(y|M_j)$
denote the prior probability and marginal likelihood of model~$M_j$,
respectively. To compare any two models, say, $M_i$ and~$M_j$, consider
the posterior odds in favor of model~$M_i$:
\[
\frac{p(M_i|y)}{p(M_j|y)}=\frac{p(M_i)p(y|M_i)}{p(M_j)p(y|M_j)}.
\]
The ratio of the marginal likelihoods, $\frac{p(y|M_i)}{p(y|M_j)}$, is
the Bayes factor and can be considered as the strength of evidence
provided by the data in favor of model $M_i$ over $M_j$. Therefore,
model comparison can be performed using marginal likelihoods once a
prior has been specified on the models. See \citet{OHaFor04}
for a review of Bayes factors and alternative methods for Bayesian
model choice. In Section~\ref{sec6.4}, we demonstrate how the variational lower
bound, a~by-product of Algorithm~\ref{algor3}, can be used in place of the log
marginal likelihood to obtain approximate posterior model
probabilities, assuming all models considered are equally probable.
Formerly, \citet{CorBis01} verified through experiments and
comparisons with cross-validation that the variational lower bound is a
good score for model selection in Gaussian mixture models.

We note that standard model selection criteria such as AIC or BIC are
difficult to apply to GLMMs, as it is not straightforward to determine
the degrees of freedom of a GLMM. \citet{YuYau12} developed a
conditional Akaike information criterion for GLMMs which takes into
account estimation uncertainty in variance component parameters.
\citet{OveFor10} considered a default strategy for Bayesian
model selection addressing issues of prior specification and
computation. See also \citet{autokey9} for a review of variable
selection methods for GLMMs.\vadjust{\goodbreak}

\section{Examples}\label{sec6}
We investigate the performance of Algorithm~\ref{algor3} using different
parametrizations by considering a simulation study and some real data
sets. When using partial noncentering, we can either initialize the
tuning parameters, $W_i$ for $i=1,\ldots,n$, and keep them as fixed or
update them at the beginning of each cycle (see Algorithm~\ref{algor3}, step 1).
Such updates are particularly useful when a good initialization is
lacking. We present results for both cases. There might not be
significant improvement in updating $W_i$ in the examples below, as the
initialization using penalized quasi-likelihood is already good.

We assessed the performance of Algorithm~\ref{algor3} using different
parametrizations by using MCMC as a ``gold standard.'' Fitting via MCMC
was performed in WinBUGS (\cite{Lun}) through \texttt{R} by using
\texttt{R2WinBUGS} (\cite{StuLigGel05}) as an interface. WinBUGS
automatically implements a Markov chain simulation for the posterior
distribution after the user specifies a model and starting values (see,
e.g., \cite{Geletal04}). We used the centered parametrization when
specifying the model in \mbox{WinBUGS}, as this produced better mixing than
the noncentered parametrization for most of the examples considered
(see \cite{BroZho10}). The MCMC algorithm was initialized similarly
using the fit from penalized quasi-likelihood. In each case, three
chains were run simultaneously to assess convergence, each with 50,000
iterations, and the first 5000 iterations were discarded in each chain
as burn-in. A thinning factor of 10 was applied to reduce dependence
between draws. The posterior means and standard deviations reported
were based on the remaining 13,500 iterations. The computation times
reported for MCMC are the times taken for updating in WinBUGS. We used
the same priors for MCMC and Algorithm~\ref{algor3}. For the fixed
effects, we used a $N(0, 1000I)$ prior. All code was written in the R
language and run on a dual processor Windows PC 3.30 GHz workstation.

\subsection{Simulated Data}\label{sec6.1}
In this simulation study we consider the Poisson random intercept model
\[
y_{ij}|u_i \sim\operatorname{Poisson} \bigl( \exp(
\beta_0+\beta_1 x_{ij}+u_i) \bigr)
\]
and the logistic random intercept model
\[
y_{ij}|u_i \sim\operatorname{Bernoulli} \biggl(
\frac{\exp(\beta_0+\beta
_1 x_{ij}+u_i)}{1+\exp(\beta_0+\beta_1 x_{ij}+u_i)} \biggr),
\]
where $u_i \sim N(0,\sigma^2)$. For the Poisson random intercept model,
we set\vadjust{\goodbreak} $x_{ij}=j-1$ for $i=1,\ldots,100$, $j=1,2$, and used $\beta
_0=\beta _1=-0.5$, $\sigma=0.1$. For the logistic random intercept
model, we set $x_{ij}=\frac{j}{8}$, for $i=1,\ldots,50$,
$j=1,\ldots,8$, and used $\beta _0=0$, $\beta_1=5$,
$\sigma=\sqrt{1.5}$. Similar settings have been considered by
\citet{OrmWan12}. For each model, 100 data sets were generated. No
convergence issues were encountered for these simulated data, but
experience with other simulated data sets (not shown) indicate that
problems may arise when the covariance matrix of the fixed effects
estimated from penalized quasi-likelihood is nearly singular or when
the standard deviation of the random effects are very close to zero. In
such cases, we can use alternative means of initialization such as
estimates from the generalized linear model obtained by setting the
random effects as zero. The expression in (\ref{IWprior}) can also
serve as a prior guess for $D$ (see \cite{KasNat06}). Table
\ref{simstudy} reports the estimates from penalized quasi-likelihood
and the posterior means and standard deviations estimated by Algorithm
\ref{algor3} (using different parametrizations) and MCMC. Results are
averaged over the 100 sets of simulated data. We have also included
root mean squared errors computed as $\sqrt{\frac
{1}{100}\sum_{l=1}^{100} (\hat{\vartheta}_l-\vartheta_l^0)^2}$ for an
estimate $\hat{\vartheta}_l$ from the $l$th simulated data set obtained
from penalized quasi-likelihood or Algorithm~\ref{algor3} where
$\vartheta_l^0$ is the corresponding estimate from MCMC regarded as the
``gold standard.''

\begin{table*}
\def\arraystretch{0.9}
\tabcolsep=0pt
\caption{Results of simulation study showing
initialization values from penalized quasi-likelihood, posterior means
and standard deviations estimated by Algorithm \protect\ref{algor3} (different
parametrizations) and MCMC, computation times (seconds) and variational
lower bounds ($\mathcal{L}$), averaged over 100 sets of simulated data.
Values in () are the corresponding root mean squared errors}
\label{simstudy}
\begin{tabular*}{\tablewidth}{@{\extracolsep{\fill}}lclllllld{3.1}c@{}}
\hline
\textbf{Model} & \textbf{Method} & \multicolumn{1}{c}{$\bolds{\beta_0}$}
& \multicolumn{1}{c}{$\bolds{\operatorname{SE}(\beta_0)}$}
& \multicolumn{1}{c}{$\bolds{\beta_1}$} & \multicolumn{1}{c}{$\bolds{\operatorname{SE}(\beta_1)}$}
& \multicolumn{1}{c}{$\bolds{\sigma}$}
& \multicolumn{1}{c}{$\bolds{\operatorname{SE}(\sigma)}$}
& \multicolumn{1}{c}{\textbf{Time}} & \multicolumn{1}{c@{}}{$\bolds{\mathcal{L}}$} \\ \hline
Poisson & Penalized  & $-$0.54 (0.11) &
0.13 (0.02) & $-$0.48 (0.01) & 0.19 (0.03) & 0.27 (0.35) & \multicolumn{1}{c}{---}&
0.1 & \multicolumn{1}{c@{}}{---}\\
& quasi-likelihood\\
&Noncentered & $-$0.63 (0.01) & 0.13 (0.02) & $-$0.49 ($<$0.005) & 0.21
($<$0.005) & 0.48 (0.02)& 0.03 (0.08) & 3.6 & $-$196.0 \\
&Centered & $-$0.63 (0.01) & 0.05 (0.10) & $-$0.50 (0.01) & 0.16 (0.05) &
0.50 (0.01) & 0.04 (0.07) & 4.3 & $-$197.0 \\
&Partially & $-$0.63 (0.01) & 0.13 (0.02) &
$-$0.49 ($<$0.005) & 0.20 (0.01) & 0.49 (0.01) & 0.03 (0.08) & 3.5 &
$-$196.0\\
& noncentered: \\
& $W_i$ fixed\\
&Partially  & $-$0.63 (0.01) & 0.13 (0.02) &
$-$0.49 ($<$0.005) & 0.19 (0.02) & 0.49 (0.01) & 0.03 (0.08) & 4.0 &
$-$196.0\\
& noncentered: \\
& $W_i$ updated\\
&MCMC & $-$0.64 & 0.15 & $-$0.48 & 0.21 & 0.50 & 0.11 & 60.1 & \multicolumn{1}{c@{}}{---}
\\ [6pt]
Logistic & Penalized & $-$0.10 (0.06)
& 0.32 (0.07) & \hphantom{$-$}5.02 (0.27) & 0.63 (0.24) & 1.25 (0.16) & \multicolumn{1}{c}{---}&
0.2 & \multicolumn{1}{c@{}}{---}\\
& quasi-likelihood\\
& Noncentered & $-$0.07 (0.02) & 0.33 (0.06) & \hphantom{$-$}5.20 (0.04) & 0.77 (0.09)
& 1.18 (0.06)& 0.12 (0.20) & 3.2 & $-$140.4 \\
& Centered & $-$0.07 (0.02) & 0.17 (0.21) & \hphantom{$-$}5.24 (0.02) & 0.41 (0.45) &
1.24 (0.03)& 0.13 (0.20) & 3.1 & $-$141.1 \\
& Partially & $-$0.07 (0.02) & 0.30 (0.09) &
\hphantom{$-$}5.23 (0.02) & 0.50 (0.37) & 1.22 (0.03)& 0.12 (0.20) & 2.9 & $-$140.5 \\
& noncentered:\\
& $W_i$ fixed\\
& Partially  & $-$0.07 (0.02) & 0.30 (0.08) &
\hphantom{$-$}5.21 (0.04) & 0.50 (0.36) & 1.22 (0.04)& 0.12 (0.20) & 3.9 & $-$140.5\\
& noncentered:\\
& $W_i$ updated \\
& MCMC & $-$0.05 & 0.38 & \hphantom{$-$}5.23 & 0.85 & 1.24 & 0.32 & 146.6 & \multicolumn{1}{c@{}}{---}
\\
\hline
\end{tabular*}
\end{table*}

For the Poisson model, the posterior means of the fixed effects and
random effects estimated using the centered and noncentered
parametrizations are quite close and also close to that of MCMC.
However, the posterior standard deviations of the fixed effects are
underestimated in the centered parame\-trization and the noncentered
parametrization does better. The average time to convergence was
shorter with noncentering and a higher lower bound was attained on
average. We observe that the partially noncentered parametrization
where tuning parameters were not updated took on average the least time
to converge and produced a fit closer to that of the noncentered
parametrization but with improvements in the estimation of the
posterior means of the random effects. When the tuning parameters were
updated, the fit was just as good, although computation time was
longer. For the logistic model, centering and noncentering have
different merits. While centering produced better estimates of the
posterior means, the posterior standard deviations of the fixed effects
were underestimated. The partially noncentered\vadjust{\goodbreak} parametrization tries to
adapt between the centered and noncentered parametrizations, producing
better estimates of the posterior means than noncentering and better
estimates of the posterior standard deviations than centering. When the
tuning parameters were updated, the results leaned more toward the
noncentered parame\-trization and the algorithm took longer to converge.
In both cases, Algorithm~\ref{algor3} using the partially noncentered
parametrization was faster than MCMC and provided better estimates of
the fixed effects and random effects than penalized quasi-likelihood.
There are some difficulties, however, in comparing Algorithm~\ref{algor3} and MCMC
in this way, as the time taken for Algorithm~\ref{algor3} to converge depends on
the initialization, stopping rule and the rate of convergence also
depends on the problem. Similarly, the updating time taken for MCMC is
also problem-dependent and depends on the length of burn-in and number
of sampling iterations. In addition, we observed (in simulated data
sets not shown) that posterior inferences can be sensitive to prior
assumptions on the variance components in Poisson models where many of
the counts are close to zero or in binary data where the cluster size
is small (see \cite{BroDra06} and \cite{RooHel11}).\vadjust{\goodbreak}

\subsection{Epilepsy Data}\label{sec6.2}
Here we consider the epilepsy data of \citet{ThaVai90} which has
been analyzed by many authors (see, e.g., \cite{BreCla93}; \cite{OrmWan12}).
In this clinical trial, 59 epileptics were randomized to a new
anti-epileptic drug, progabide ($\mathrm{Trt}=1$) or a placebo ($\mathrm{Trt}=0$). Before
receiving treatment, baseline data on the number of epileptic seizures
during the preceding 8-week period were recorded. The logarithm of
$\frac{1}{4}$ the number of baseline seizures (Base) and the logarithm
of age (Age) were treated as covariates. Counts of epileptic seizures
during the 2 weeks before each of four successive clinic visits (Visit,
%
\begin{table*}
\caption{Results for epilepsy data models II and IV
showing initialization values from penalized quasi-likelihood,\break
posterior means and standard deviations (values after $\pm$) estimated
by Algorithm \protect\ref{algor3} (different parametrizations)\break and MCMC, computation times
(seconds) and variational lower bounds ($\mathcal{L}$)}\label{modelII}
\begin{tabular*}{\tablewidth}{@{\extracolsep{\fill}}lcccccc@{}}
\hline
& \multicolumn{1}{c}{\textbf{Penalized}} &  &  & \multicolumn{1}{c}{\textbf{Partially}}
& \multicolumn{1}{c}{\textbf{Partially}}  &
\\
& \multicolumn{1}{c}{\textbf{quasi-}} & &
&\multicolumn{1}{c}{\textbf{noncentered:}}
& \multicolumn{1}{c}{\textbf{noncentered:}} \\
& \multicolumn{1}{c}{\textbf{likelihood}} & \multicolumn{1}{c}{\textbf{Noncentered}}
& \multicolumn{1}{c}{\textbf{Centered}} &
\multicolumn{1}{c}{$\bolds{W_i}$ \textbf{fixed}}
& \multicolumn{1}{c}{$\bolds{W_i}$ \textbf{updated}} & \multicolumn{1}{c@{}}{\textbf{MCMC}}\\
\hline
Model II\\
\quad$\beta_0$ & $0.31 \pm 0.26$ & $0.26 \pm 0.11$ & $0.27 \pm 0.24$ & $0.27
\pm 0.26$ & $0.27 \pm 0.27$ & $0.26 \pm 0.27$ \\
\quad$\beta_{\mathrm{Base}}$ & $0.88 \pm 0.13$ & $0.89 \pm 0.04$ & $0.88 \pm
 0.13$ & $0.88 \pm 0.13$ & $0.88 \pm 0.14$ & $0.89 \pm 0.14$ \\
\quad$\beta_{\mathrm{Trt}}$ & $-0.91 \pm 0.41$ & $-0.94 \pm 0.15$ & $-0.94
\pm 0.36$ & $-0.94 \pm 0.40$ & $-0.94 \pm 0.41$ & $-0.94 \pm 0.42$ \\
\quad$\beta_{\mathrm{Base} \times\mathrm{Trt}}$ & $0.34 \pm 0.20$ & $0.34
\pm 0.06$ & $0.34 \pm 0.19$ & $0.34 \pm 0.21$ &$0.34 \pm 0.21$ & $0.34
\pm 0.21$ \\
\quad$\beta_{\mathrm{Age}}$ & $0.54 \pm 0.35$ & $0.50 \pm 0.12$ & $0.48 \pm
0.33$ & $0.48 \pm 0.35$ & $0.48 \pm 0.36$ & $0.48 \pm 0.37$ \\
\quad$\beta_{\mathrm{V4}}$& $-0.16 \pm 0.08$ & $-0.16 \pm 0.05$ & $-0.16 \pm
 0.05$ &$-0.16 \pm 0.05$ & $-0.16 \pm 0.05$ & $-0.16 \pm 0.05$ \\
\quad$\sigma$ & 0.44 & $0.50 \pm 0.05$ & $0.54 \pm 0.05$ & $0.53 \pm 0.05$&
$0.53 \pm 0.05$ & $0.53 \pm 0.06$\\
\quad$\mathcal{L}$ & ---&$-707.3$ & $-702.0$ & $-701.6$ & $-701.5$ &
---\\
\quad Time &0.2 & 1.1 & 0.4 & 0.4 & 0.6 & 61 \\
[6pt]
Model IV\\
\quad$\beta_0$ & $0.27 \pm 0.26$ & $0.21 \pm 0.10$ & $0.21 \pm 0.24$ & $0.21
\pm 0.26$ & $0.21 \pm 0.26$ & $0.21 \pm 0.27$ \\
\quad$\beta_{\mathrm{Base}}$ & $0.88 \pm 0.13$ & $0.89 \pm 0.04$ & $0.88 \pm
 0.13$ & $0.89 \pm 0.13$ & $0.89 \pm 0.13$ & $0.88 \pm 0.14$ \\
\quad$\beta_{\mathrm{Trt}}$ & $-0.92 \pm 0.41$ & $-0.94 \pm 0.15$ & $-0.93
\pm 0.36$ & $-0.93 \pm 0.40$ & $-0.93 \pm 0.40$ & $-0.94 \pm 0.42$ \\
\quad$\beta_{\mathrm{Base} \times\mathrm{Trt}}$ & $0.35 \pm 0.20$ & $0.34
\pm 0.06$ & $0.34 \pm 0.19$ & $0.34 \pm 0.20$ &$0.34 \pm 0.21$ & $0.34
\pm 0.22$ \\
\quad$\beta_{\mathrm{Age}}$ & $0.54 \pm 0.35$ & $0.49 \pm 0.12$ & $0.47 \pm
0.32$ & $0.47 \pm 0.35$ & $0.47 \pm 0.35$ & $0.47 \pm 0.37$ \\
\quad$\beta_{\mathrm{Visit}}$& $-0.28 \pm 0.16$ & $-0.27 \pm 0.10$ & $-0.27
\pm 0.10$ &$-0.27 \pm 0.14$ & $-0.27 \pm 0.15$ & $-0.27 \pm 0.17$ \\
\quad$\sigma_{11}$ & 0.45 & $0.50 \pm 0.05$ & $0.53 \pm 0.05$ & $0.52 \pm
0.05$ & $0.53 \pm 0.05$ & $0.53 \pm 0.06$ \\
\quad$\sigma_{22}$ & 0.46 & $0.75 \pm 0.07$ & $0.77 \pm 0.07$ & $0.75 \pm
0.07$ & $0.76 \pm 0.07$ & $0.76 \pm 0.15$ \\
\quad$\mathcal{L}$ & ---&$-701.4$ & $-696.1$ & $-695.3$ & $-695.1$ &
---\\
\quad Time &0.5 & 1.5 & 1.3 & 1.2 & 1.4 & 122 \\
\hline
\end{tabular*}
\end{table*}
coded as $
\mathrm{Visit}_1=-0.3$, $\mathrm{Visit}_2=-0.1$, $\mathrm{Visit}_3=0.1$
and $\mathrm{Visit}_4=0.3$) were recorded. A binary variable ($\mathrm{V4}=1$ for
fourth visit, 0 otherwise) was also considered as a covariate. We
consider models II and~IV from \citet{BreCla93}. Model II is
a Poisson random intercept model where
\begin{eqnarray*}
\log\mu_{ij}&=& \beta_0+\beta_{\mathrm{Base}}
\mathrm{Base}_i+\beta _{\mathrm{Trt}} \mathrm{Trt}_i\\
&&{}+
\beta_{\mathrm{Base} \times\mathrm
{Trt}} \mathrm{Base}_i \times\mathrm{Trt}_i+
\beta_{\mathrm{Age}} \mathrm{Age}_i\\
&&{} +\beta_{\mathrm{V4}}
\mathrm{V4}_{ij}+u_i
\end{eqnarray*}
for $i=1,\ldots,n$, $j=1,\ldots,4$ and $u_i\sim N(0,\sigma^2)$. Mod\-el~IV is a
Poisson random intercept and slope model of the form
\begin{eqnarray*}
\log\mu_{ij}&=& \beta_0+\beta_{\mathrm{Base}}
\mathrm{Base}_i+\beta _{\mathrm{Trt}} \mathrm{Trt}_i\\
&&{}+
\beta_{\mathrm{Base} \times\mathrm
{Trt}} \mathrm{Base}_i \times\mathrm{Trt}_i+
\beta_{\mathrm{Age}} \mathrm{Age}_i \\
&&{}+\beta_{\mathrm{Visit}}
\mathrm{Visit}_{ij}
+u_{1i}+u_{2i} \mathrm{Visit}_{ij}
\end{eqnarray*}
for $i=1,\ldots,n$, $j=1,\ldots,4$ and
\[
\left[\matrix{ u_{1i}
\cr
u_{2i}} \right]\sim N\lleft(0,\left[\matrix{ \sigma_{11}^2 &
\sigma_{12}
\cr
\sigma_{21} & \sigma_{22}^2}
\right] \rright).
\]
As the MCMC chains for intercept and
Age were mixing poorly, we decided to center the covariate Age. In the
analysis that follows, we assume $\mathrm{Age}_i$ has been replaced by
$\mathrm{Age}_i-\operatorname{mean}(\mathrm{Age})$.

Table~\ref{modelII} shows the estimates of the posterior means and
standard deviations of the fits from MCMC and Algorithm~\ref{algor3} (using
different parametrizations), initialization values from penalized
quasi-likelihood and computation times in seconds taken by different
methods. All the variational methods are faster than MCMC by an order
of magnitude which is especially important in\vadjust{\goodbreak} large scale applications.
In the noncentered parametrization, the standard deviations of the
fixed effects were underestimated and the centered parametrization does
better in this aspect. The partially noncentered parametrization
produced a fit that is closer to that of the centered parametrization
and improved upon it. In both models, the fits produced by partial
noncentering are very close to that produced by MCMC and are superior
to that of the centered and noncentered parametrizations. The lower
bound attained by partial noncentering is also higher than that of
centering and noncentering, giving a tighter bound on the log marginal
likelihood. It is important to emphasize that the relevant comparison
\setcounter{figure}{1}
\begin{figure*}

\includegraphics{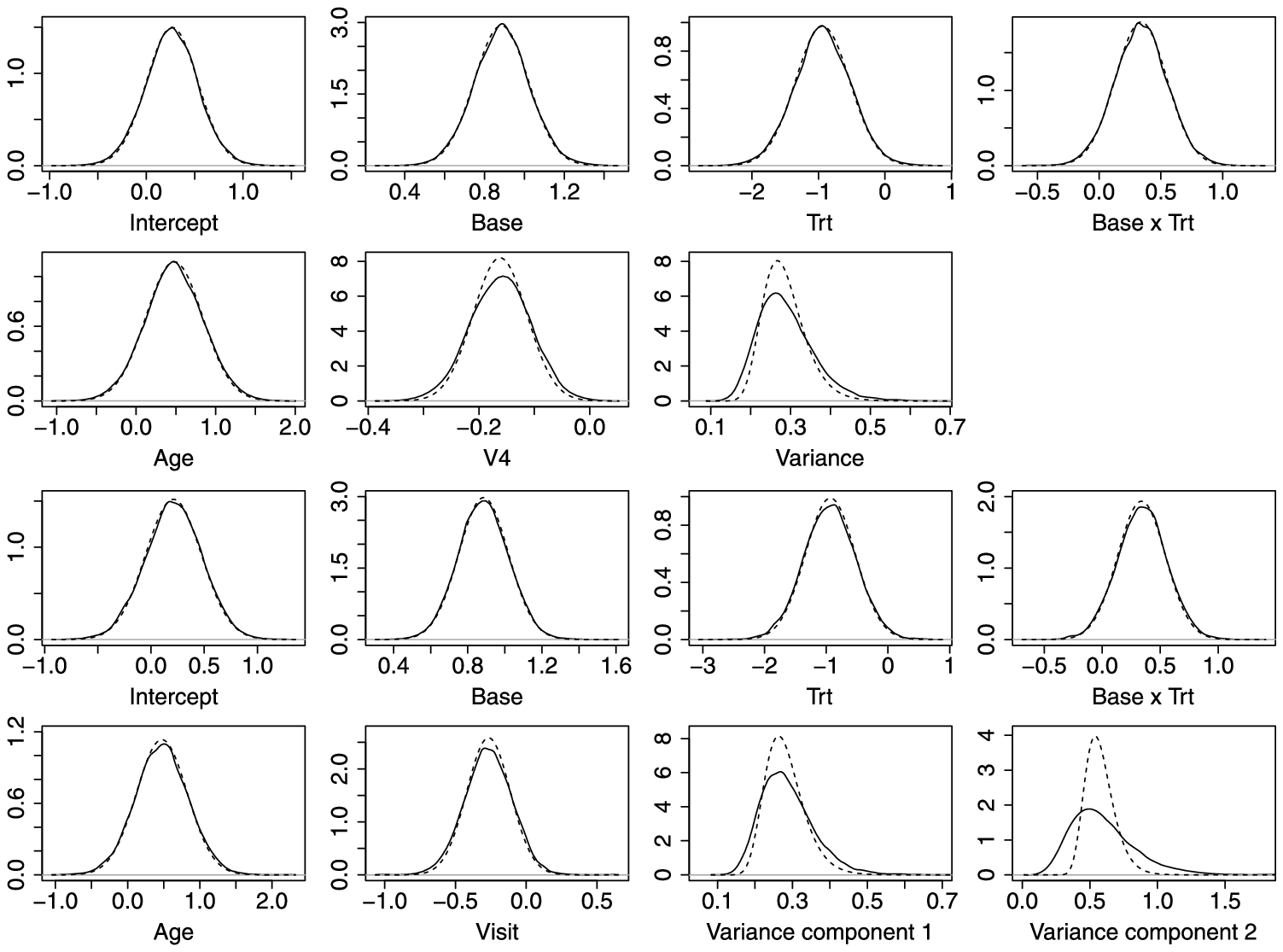}

\caption{Marginal posterior distributions for parameters in model II
(first two rows) and model IV (last two rows) of the epilepsy data
estimated by MCMC (solid line) and Algorithm \protect\ref{algor3} using partially
noncentered parametrization where tuning parameters are updated (dashed line).}
\label{modelIIIVfig}
\end{figure*}
is of the partially noncentered parametrization to the worst of the
centered and noncentered parametrizations, since in general we do not
know if centering or noncentering is better without running both
algorithms. Partial noncentering, on the other hand, automatically
chooses a near optimal parametrization. Updating of the tuning
parameters helped to improve the fit produced by partial noncentering.
Figure~\ref{modelIIIVfig} shows the marginal posterior distributions
for parameters in models II and IV estimated by MCMC (solid line) and
Algorithm~\ref{algor3} using the partially noncentered parametrization where
tuning parameters are updated (dashed line). The variational posterior
densities of the fixed effects are very close to those obtained via
MCMC. For the variance components, there is still some underestimation
of the posterior variance.

\subsection{Toenail Data}\label{sec6.3}
This data set was obtained from a multicenter study comparing two
competing oral antifungal treatments for toenail infection
(\cite{DeBetal98}), courtesy of Novoartis, Belgium. It contains
information for 294~patients to be evaluated at seven visits. Not all
patients attended all seven planned visits and there were 1908
measurements in total. The patients were randomized into two treatment
groups, one group receiving 250 mg per day of terbinafine ($\mathrm{Trt}=1$) and
the other group 200 mg per day of itraconazole ($\mathrm{Trt}=0$). Visits were
planned at weeks 0, 4, 8, 12, 24, 36 and~48, but patients did not
always arrive as scheduled\vadjust{\goodbreak} and the exact time in months ($t$) that they
did attend was recorded. The binary response variable (onycholysis)
indicates the degree of separation of the nail plate from the nail bed
(0 if none or mild, 1 if moderate or severe). We consider the following
logistic random intercept model,
\[
\operatorname{logit}(\mu_{ij}) = \beta_0+
\beta_{\mathrm{Trt}} \mathrm {Trt}_i+\beta_t
t_{ij} +\beta_{\mathrm{Trt} \times t} \mathrm{Trt}_i \times
t_{ij}+u_i,
\]
where $u_i\sim N(0,\sigma^2)$ for $i=1,\ldots,294$, $1\leq j \leq7$.

%
\begin{table*}
\caption{Results for toenail data showing values
used for initialization from penalized quasi-likelihood,\break posterior
means and posterior standard deviations (values after $\pm$) from
Algorithm \protect\ref{algor3} (different parametrizations)\break and MCMC, computation times
(seconds) and variational lower bounds ($\mathcal{L}$)}
\label{toenailtab}
\begin{tabular*}{\tablewidth}{@{\extracolsep{\fill}}lcccccc@{}}
\hline
& \multicolumn{1}{c}{\textbf{Penalized}} &  &  & \multicolumn{1}{c}{\textbf{Partially}}
& \multicolumn{1}{c}{\textbf{Partially}}  &
\\
& \multicolumn{1}{c}{\textbf{quasi-}} & &
&\multicolumn{1}{c}{\textbf{noncentered:}}
& \multicolumn{1}{c}{\textbf{noncentered:}} \\
& \multicolumn{1}{c}{\textbf{likelihood}} & \multicolumn{1}{c}{\textbf{Noncentered}}
& \multicolumn{1}{c}{\textbf{Centered}} &
\multicolumn{1}{c}{$\bolds{W_i}$ \textbf{fixed}}
& \multicolumn{1}{c}{$\bolds{W_i}$ \textbf{updated}} & \multicolumn{1}{c@{}}{\textbf{MCMC}}\\
\hline
$\beta_0$ & $-0.75 \pm 0.25$ & $-1.41 \pm 0.17$ & $-1.44 \pm 0.29$ &
$-1.44 \pm 0.35$ & $-1.44 \pm 0.32$ & $-1.65 \pm 0.44$ \\
$\beta_{\mathrm{Trt}}$ & $-0.04 \pm 0.35$ & $-0.13 \pm 0.25$ & $-0.13
\pm 0.41$ & $-0.13 \pm 0.49$ & $-0.13 \pm 0.45$ & $-0.17 \pm 0.60$ \\
$\beta_{\mathrm{t}}$ & $-0.30 \pm 0.03$ & $-0.38 \pm 0.04$ & $-0.38 \pm
 0.03$ & $-0.38 \pm 0.03$ & $-0.38 \pm 0.03$ & $-0.40 \pm 0.05$ \\
$\beta_{\mathrm{Trt} \times\mathrm{Time}}$& $-0.10 \pm 0.05$ & $-0.13
\pm 0.06$ & $-0.13 \pm 0.04$ & $-0.13 \pm 0.04$ &$-0.13 \pm 0.04$ &
$-0.14 \pm 0.07$ \\
$\sigma$ & 2.32 & $3.52 \pm 0.15$ & $3.56 \pm 0.15$ & $3.55 \pm 0.15$ &
$3.55 \pm 0.15$ & $4.10 \pm 0.39$ \\
$\mathcal{L}$ & ---&$-664.1$ & $-663.1$ & $-662.7$ & $-662.9$ &
---\\
Time & 2.8 & 37.9 & 27.9 & 26.0 & 24.1 & 1072 \\
\hline
\end{tabular*}
\end{table*}

Table~\ref{toenailtab} shows the posterior means and standard
deviations of the fits from MCMC and Algorithm~\ref{algor3} (using different
parametrizations), initialization values from penalized
quasi-likelihood and computation time in seconds taken by different
methods. Again, the VB methods are faster than MCMC by an order of
magnitude. In this example, centering produced a better fit than
noncentering and partial noncentering produced a fit closer to that of
the centered parametrization but improving it. Partial noncentering
also took less time to converge and attained a lower bound higher than
that of the centered and noncentered parametrizations. Again, we\vadjust{\goodbreak}
emphasize that it is not easy to know beforehand which of centering or
noncentering will perform better, and a big advantage of partial
noncentering is the way that it automatically chooses a good
parametrization. In this example, updating the tuning parameters did
not result in a better fit although the time to convergence is reduced.
\setcounter{figure}{2}
\begin{figure*}[b]

\includegraphics{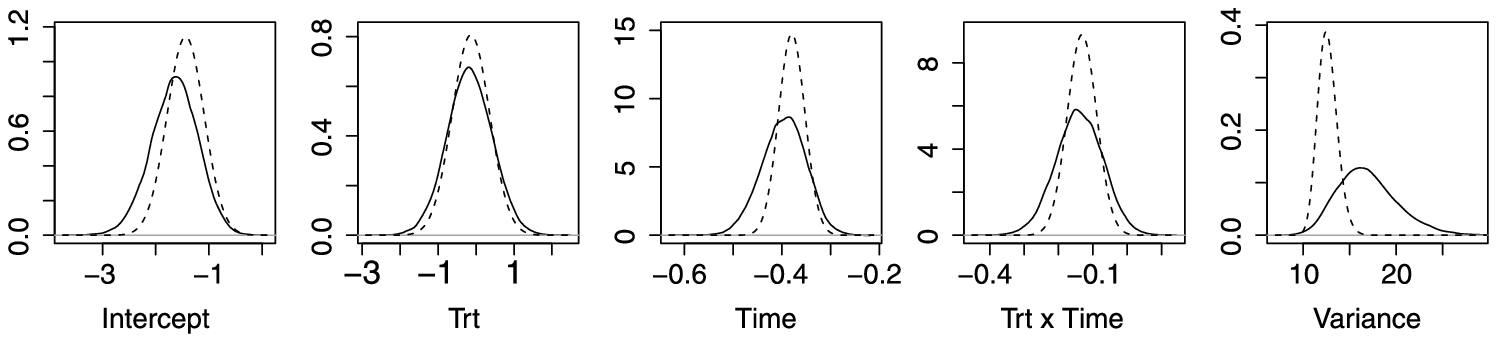}

\caption{Marginal posterior distributions for parameters in toenail
data estimated by MCMC (solid line) and Algorithm \protect\ref{algor3} using partially
noncentered parametrization (tuning parameters not updated) (dashed line).}
\label{toenailfig}
\end{figure*}
The marginal posterior distributions estimated by\break MCMC (solid line) and
Algorithm~\ref{algor3} using the partially noncentered parametrization where
tuning parameters were not updated (dashed line) are shown in Figure
\ref{toenailfig}. Compared with the MCMC fit, there is still some
underestimation of the variance of the fixed effects particularly for
the parameters which could not be centered. Although the partially
noncentered parametrization has improved the estimation of the random
effects from the initial penalized quasi-likelihood fit, there is still
some underestimation of the mean and variance of the random effects
when compared to the MCMC fit.

\subsection{Six Cities Data}\label{sec6.4}
In the previous two real data examples, centering performed better than
noncentering and partial noncentering\vadjust{\goodbreak} was able to improve on the
centering results. While centering often performs better than
noncentering, we use this example to show that partial noncentering
will automatically tend toward noncentering when noncentering is
preferred. We consider the six cities data in \citet{FitLai93},
where the binary response variable $y_{ij}$ indicates the wheezing
status (1~if wheezing, 0 if not wheezing) of the $i$th child at
time-point $j$, $i=1,\ldots,537$, $j=1,2,3,4$. We use as covariate the
age of the child at time-point $j$, centered at 9 years (Age), and
consider the following random intercept and slope model:
\[
\operatorname{logit}(\mu_{ij})=\beta_0+
\beta_{\mathrm{Age}}\mathrm {Age}_i+u_{1i}+u_{2i}
\mathrm{Age}_i
\]
for $i=1,\ldots,537$, $j=1,\ldots,4$ and
\[
\left[\matrix{ u_{1i}
\cr
u_{2i} } \right]\sim N\lleft(0,\left[\matrix{ \sigma_{11}^2 &
\sigma_{12}
\cr
\sigma_{21} & \sigma_{22}^2}
\right] \rright).
\]
This model has been considered in
\citet{OveFor10}.

\begin{table*}
\caption{Results for six cities data showing values used
for initialization from penalized quasi-likelihood,\break posterior means and
posterior standard deviations (values after $\pm$) from Algorithm \protect\ref{algor3}
(different parametrizations)\break and MCMC, computation times (seconds) and
variational lower bounds ($\mathcal{L}$)}\label{sixtab}
\begin{tabular*}{\tablewidth}{@{\extracolsep{\fill}}lcccccc@{}}
\hline
& \multicolumn{1}{c}{\textbf{Penalized}} &  &  & \multicolumn{1}{c}{\textbf{Partially}}
& \multicolumn{1}{c}{\textbf{Partially}}  &
\\
& \multicolumn{1}{c}{\textbf{quasi-}} & &
&\multicolumn{1}{c}{\textbf{noncentered:}}
& \multicolumn{1}{c}{\textbf{noncentered:}} \\
& \multicolumn{1}{c}{\textbf{likelihood}} & \multicolumn{1}{c}{\textbf{Noncentered}}
& \multicolumn{1}{c}{\textbf{Centered}} &
\multicolumn{1}{c}{$\bolds{W_i}$ \textbf{fixed}}
& \multicolumn{1}{c}{$\bolds{W_i}$ \textbf{updated}} & \multicolumn{1}{c@{}}{\textbf{MCMC}}\\
\hline
$\beta_0$ & $-3.12 \pm 0.14$ & $-3.05 \pm 0.09$ & $-3.05 \pm 0.09$ &
$-3.05 \pm 0.13$ & $-3.05 \pm 0.13$ & $-3.29 \pm 0.25$ \\
$\beta_{\mathrm{Age}}$ & $-0.24 \pm 0.08$ & $-0.22 \pm 0.07$ & $-0.21
\pm 0.02$ & $-0.22 \pm 0.07$ & $-0.22 \pm 0.07$ & $-0.25 \pm 0.16$ \\
$\sigma_{11}$ & 2.52 & $2.16 \pm 0.07$ & $2.16 \pm 0.07$ & $2.16 \pm
0.07$ & $2.16 \pm 0.07$ & $2.48 \pm 0.24$ \\
$\sigma_{22}$ & 1.19 & $0.55 \pm 0.02$ & $0.56 \pm 0.02$ & $0.55 \pm
0.02$ & $0.55 \pm 0.02$ & $0.61 \pm 0.10$ \\
$\mathcal{L}$ & ---&$-833.2$ & $-834.1$ & $-832.8$ & $-832.6$ &
---\\
Time & 3.8 & 114.7 & 125.8 & 110.6 & 120.6 & 1010 \\
\hline
\end{tabular*}
\end{table*}

Table~\ref{sixtab} shows the estimates of the posterior means and
standard deviations of the fits from MCMC and Algorithm~\ref{algor3} using
different parametrizations, the values\vadjust{\goodbreak} from penalized quasi-likelihood
used for initialization and the computation times in seconds taken by
different methods. Noncentering performed better than centering in this
case with a shorter time to convergence, higher lower bound and a
better estimate of the posterior standard deviation of
$\beta_{\mathrm{Age}}$. Partial noncentering further improved upon the
results of noncentering with an improved estimate of the posterior
standard deviation of $\beta_0$ and faster convergence. All the
variational methods are again faster than MCMC by an order of
magnitude.\looseness=1

\subsection{Owl Data}\label{sec6.5}
In this example we illustrate the use of the variational lower bound, a
by-product of Algorithm~\ref{algor3}, for model selection. For MCMC, on the other
hand, it is not straightforward in general to get a good estimate of
the marginal likelihood based on the MCMC output. It is also not always
obvious how to apply standard model selection criteria like AIC and BIC
to hierarchical models like GLMMs.

\citet{RouBer07} analyzed the begging behavior of nestling
barn owls and looked at whether offspring beg for food at different
intensities from the mother than father. They sampled $n=27$ nests and
counted the number of calls made by all offspring in the absence of
parents. Half of the nests were given extra prey, and from the other
half prey were removed. Measurements took place on two\break nights, and food
treatment was swapped the second night. The number of measurements at
each nest ranged from 4 to 52 with a total of 599. We use as covariates
sex of parent ($\mathrm{Sex}=1$ if male, 0 if female), the time at which a parent
arrived with a prey ($t$), and food treatment ($\mathrm{Trt} = 1$ if ``satiated,'' 0
if ``deprived''). The number of nestlings per nest (broodsize, $E$)
ranged from 1 to 7.\vadjust{\goodbreak}

%
\begin{table*}
\caption{Variational lower bounds for owl data models 1
to 11 and computation time in brackets}\label{owltab}
\begin{tabular*}{\tablewidth}{@{\extracolsep{\fill}}lllll@{}}
\hline
&&& \multicolumn{1}{c}{\textbf{Partially}} & \multicolumn{1}{c@{}}{\textbf{Partially}}\\
&&& \multicolumn{1}{c}{\textbf{noncentered:}} & \multicolumn{1}{c@{}}{\textbf{noncentered:}}\\
& \multicolumn{1}{c}{\textbf{Noncentered}} & \multicolumn{1}{c}{\textbf{Centered}}
&  \multicolumn{1}{c}{$\bolds{W_i}$ \textbf{fixed}} &
\multicolumn{1}{c@{}}{$\bolds{W_i}$ \textbf{updated}} \\
\hline
First stage\\
\quad Model 1& $-$2544.6 (0.2) & $-$2543.7 (0.3) & $-$2543.6 (0.4) & $-$2543.7 (0.6)
\\
\quad Model 2& $-$2537.6 (0.2) & $-$2536.6 (0.3) & $-$2536.6 (0.4) & $-$2536.6 (0.5)
\\
\quad Model 3& $-$2540.2 (0.2) & $-$2539.2 (0.3) & $-$2539.2 (0.3) & $-$2539.2 (0.5)
\\
\quad Model 4& $-$2533.2 (0.2) & $-$2532.1 (0.3) & $-$2532.1 (0.3) & $-$2532.1 (0.4)
\\ [6pt]
Second stage\\
\quad Model 5& $-$2527.0 (0.2) & $-$2525.5 (0.2) & $-$2525.5 (0.2) & $-$2525.4 (0.3)
\\
\quad Model 6& $-$2628.3 (0.2) & $-$2627.2 (0.3) & $-$2627.1 (0.3) & $-$2627.1 (0.5)
\\
\quad Model 7& $-$2664.0 (0.2) & $-$2662.9 (0.2) & $-$2662.8 (0.3) & $-$2662.8 (0.4)
\\ [6pt]
Third stage\\
\quad Model 8& $-$2621.5 (0.2) & $-$2620.0 (0.2) & $-$2620.0 (0.2) & $-$2620.0 (0.3)
\\
\quad Model 9& $-$2660.4 (0.2) & $-$2658.8 (0.2) & $-$2658.8 (0.2) & $-$2658.8 (0.2)
\\
\quad Model 10& $-$2689.4 ($<$0.05) \\[6pt]
Final stage \\
\quad Model 11& $-$2448.7 (1.1) & $-$2445.7 (0.4) & $-$2445.8 (0.3) & $-$2445.6 (0.4)
\\
\hline
\end{tabular*}
\end{table*}

\citet{Zuuetal09} modeled the number of calls at nest $i$ for
the $j$th observation as a Poisson distribution with mean $\mu_{ij}$
and used log transformed broodsize as an offset with nest as a random
effect. The prime aim of their analysis was to find a sex effect and
the largest model they considered was the following:
\begin{enumerate}[10.]
\item$\log(\mu_{ij})=\log(E_{ij})+\beta_0+\beta_{\mathrm{Sex}} \mathrm
{Sex}_{ij} +\beta_{\mathrm{Trt}} \mathrm{Trt}_{ij}+\beta_t t_{ij}+\beta
_{\mathrm{Sex}\times\mathrm{Trt}}\mathrm{Sex}_{ij} \times\mathrm
{Trt}_{ij}+\beta_{\mathrm{Sex}\times t}
\mathrm{Sex}_{ij} \times t_{ij}+u_i$,
\end{enumerate}
where $\log(E_{ij})$ is an offset and $u_i\sim N(0,\sigma^2)$ for
$i=1,\ldots,27$, $j=1,\ldots,n_i$. At the recommendation of
\citet{Zuuetal09}, we center $t$ to reduce correlation of $t$ with
the intercept. Henceforth, we assume $t_{ij}$ has been replaced by
$t_{ij}- \operatorname{mean}(t)$. In the first stage, we consider
models 1 to 4 and determine if the two interaction terms should be
retained. Models 2 to 4 are as follows:
\begin{enumerate}[10.]
\item[2.]$\log(\mu_{ij})=\log(E_{ij})+\beta_0+\beta_{\mathrm{Sex}} \mathrm
{Sex}_{ij} +\beta_{\mathrm{Trt}} \mathrm{Trt}_{ij}+\beta_t t_{ij} +
\beta_{\mathrm{Sex} \times\mathrm{Trt}} \mathrm{Sex}_{ij} \times
\mathrm{Trt}_{ij}+u_i$,
\item[3.]$\log(\mu_{ij})=\log(E_{ij})+\beta_0+\beta_{\mathrm{Sex}} \mathrm
{Sex}_{ij} +\beta_{\mathrm{Trt}} \mathrm{Trt}_{ij}+\beta_t t_{ij} +
\beta_{\mathrm{Sex} \times t} \mathrm{Sex}_{ij} \times t_{ij}+u_i$,
\item[4.]$\log(\mu_{ij})=\log(E_{ij})+\beta_0+\beta_{\mathrm{Sex}} \mathrm
{Sex}_{ij} +\beta_{\mathrm{Trt}} \mathrm{Trt}_{ij}+\beta_t t_{ij}+u_i$.
\end{enumerate}
From Table~\ref{owltab}, the preferred model (with the highest lower
bound) is model 4 where both interaction terms have been dropped from
model 1. Next, we consider models 5 to 7 where the main terms sex, food
treatment and arrival time are each dropped in turn:
\begin{enumerate}[10.]
\item[5.]$\log(\mu_{ij})=\log(E_{ij})+\beta_0+\beta_{\mathrm{Trt}} \mathrm
{Trt}_{ij}+\beta_t t_{ij}+u_i$,
\item[6.]$\log(\mu_{ij})=\log(E_{ij})+\beta_0+\beta_{\mathrm{Trt}} \mathrm
{Trt}_{ij}+\beta_{\mathrm{Sex}} \mathrm{Sex}_{ij}+u_i$,
\item[7.]$\log(\mu_{ij})=\log(E_{ij})+\beta_0+\beta_t t_{ij}+\beta_{\mathrm
{Sex}} \mathrm{Sex}_{ij}+u_i$.\vadjust{\goodbreak}
\end{enumerate}
Table~\ref{owltab} indicates that model 5 is the preferred model where
the term sex of the parent has been dropped from model 4. Now we
consider dropping each of the terms food treatment and arrival time in
turn or dropping the random effects $u_i$:
\begin{enumerate}[10.]
\item[8.]$\log(\mu_{ij})=\log(E_{ij})+\beta_0+\beta_{\mathrm{Trt}} \mathrm
{Trt}_{ij}+u_i$,
\item[9.]$\log(\mu_{ij})=\log(E_{ij})+\beta_0+\beta_t t_{ij}+u_i$,
\item[10.]$\log(\mu_{ij})=\log(E_{ij})+\beta_0+\beta_{\mathrm{Trt}} \mathrm
{Trt}_{ij}+\beta_t t_{ij}$.
\end{enumerate}
Table~\ref{owltab} indicates that none of the main terms food treatment
and arrival time as well as random effects should be dropped from model
5. Finally, we consider adding a random slope for arrival time:
%
\begin{table*}
\caption{Results for owl data (model 11) showing values
used for initialization from penalized quasi-likelihood,\break posterior
means and standard deviations (values after $\pm$) from Algorithm \protect\ref{algor3}
(different parametrizations)\break and MCMC and computation times (seconds)}
\label{owltab2}
\begin{tabular*}{\tablewidth}{@{\extracolsep{\fill}}lcccccc@{}}
\hline
& \multicolumn{1}{c}{\textbf{Penalized}} &  &  & \multicolumn{1}{c}{\textbf{Partially}}
& \multicolumn{1}{c}{\textbf{Partially}}  &
\\
& \multicolumn{1}{c}{\textbf{quasi-}} & &
&\multicolumn{1}{c}{\textbf{noncentered:}}
& \multicolumn{1}{c}{\textbf{noncentered:}} \\
& \multicolumn{1}{c}{\textbf{likelihood}} & \multicolumn{1}{c}{\textbf{Noncentered}}
& \multicolumn{1}{c}{\textbf{Centered}} &
\multicolumn{1}{c}{$\bolds{W_i}$ \textbf{fixed}}
& \multicolumn{1}{c}{$\bolds{W_i}$ \textbf{updated}} & \multicolumn{1}{c@{}}{\textbf{MCMC}}\\
\hline
$\beta_0$ & $0.60 \pm 0.07$ & $0.53 \pm 0.02$ & $0.51 \pm 0.08$ & $0.51
\pm 0.08$ & $0.51 \pm 0.09$ & $0.50 \pm 0.10$ \\
$\beta_{\mathrm{Trt}}$ & $-0.55 \pm 0.08$ & $-0.57 \pm 0.03$ & $-0.57
\pm 0.03$ & $-0.57 \pm 0.03$ & $-0.57 \pm 0.03$ & $-0.57 \pm 0.04$ \\
$\beta_t$ & $-0.13 \pm 0.03$ & $-0.15 \pm 0.01$ & $-0.16 \pm 0.04$ &
$-0.16 \pm 0.04$ & $-0.16 \pm 0.04$ & $-0.16 \pm 0.05$ \\
$\sigma_{11}$ & 0.24 & $0.44 \pm 0.06$ & $0.46 \pm 0.06$ & $0.45 \pm
0.06$ & $0.46 \pm 0.06$ & $0.47 \pm 0.09$\\
$\sigma_{22}$ & 0.11 & $0.22 \pm 0.03$ & $0.23 \pm 0.03$ & $0.22 \pm
0.03$ & $0.23 \pm 0.03$ & $0.23 \pm 0.05$\\
Time & 0.4 & 1.1 & 0.4 & 0.3 & 0.4 & 255 \\
\hline
\end{tabular*}
\end{table*}
\begin{enumerate}[11.]
\item[11.]$\log(\mu_{ij})=\log(E_{ij})+\beta_0+\beta_{\mathrm{Trt}} \mathrm
{Trt}_{ij}+\beta_t t_{ij}+u_{1i}+u_{2i} t_{ij}$,
\end{enumerate}
where
\[
\left[\matrix{u_{1i}
\cr
u_{2i}} \right] \sim N\lleft(0,\left[\matrix{ \sigma_{11}^2 &
\sigma_{12}
\cr
\sigma_{21} & \sigma_{22}^2}
\right] \rright).
\]
From Table~\ref{owltab}, the optimal model is model 11. This conclusion
is similar to that of \citet{Zuuetal09} and is the same
regardless of which parametrization was used. It is thus sufficient to
consider just the partially noncentered parametrization. The
computation time taken by Algorithm~\ref{algor3} for each model fitting is very
short and makes this a convenient way of carrying out model selection
or for narrowing down the range of likely models. Further model\vadjust{\goodbreak}
comparisons can be performed using cross-validation or other approaches.

We present the estimated posterior means and standard deviations for
the optimal model in Table~\ref{owltab2}. The marginal posterior
distributions estimated by MCMC (solid line) and Algorithm~\ref{algor3} using
partially noncentered parametrization where tuning parameters are
updated (dashed line) are shown in Figure~\ref{owlfig}. In this case,
centering produced a better fit than noncentering and partial
\setcounter{figure}{3}
\begin{figure*}[b]

\includegraphics{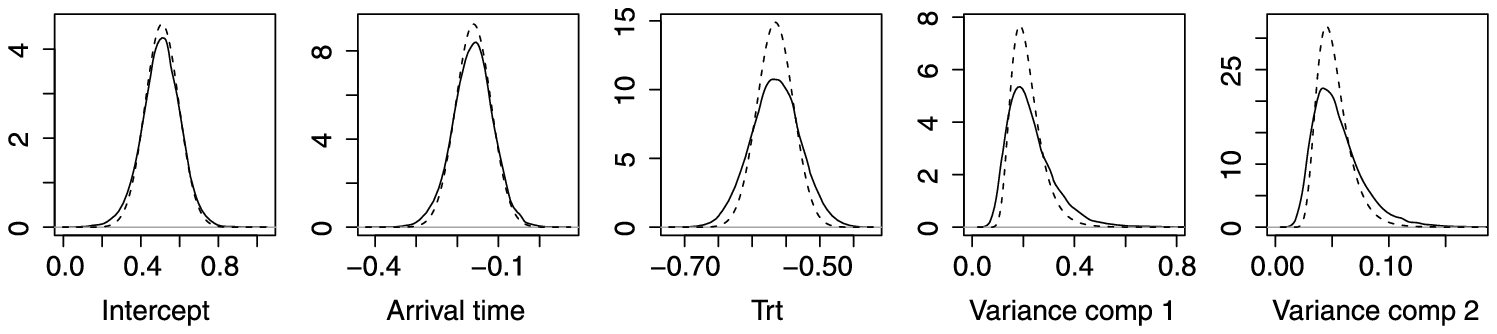}

\caption{Marginal posterior distributions for parameters in model 11
(owl data) estimated by MCMC (solid line) and Algorithm~\protect\ref{algor3} using
partially noncentered parametrization where tuning parameters are
updated (dashed line).}
\label{owlfig}
\end{figure*}
noncentering produced a fit that is close to that of centering.
Updating the tuning parameters helped to improve the fit of the
partially noncentered parametrization slightly and is closest to the
MCMC fit. From the posterior density plots, there is good estimation of
the posterior means by Algorithm~\ref{algor3} using partially noncentered
parametrization with updated tuning parameters, but there is still some
underestimation of the posterior variance.

\section{Conclusion}\label{sec7}
In this paper we described a partially noncentered parametrization for
GLMMs and compared the performance of different parametrizations using
an \mbox{algorithm} called nonconjugate variational message passing developed
recently in machine learning. Focusing on Poisson and logistic mixed
models, we applied our methods to analysis of longitudinal data sets.\vadjust{\goodbreak}
For the logistic model, some parameter updates were not available in
closed form and we used adaptive Gauss--Hermite quadrature to
approximate the intractable integrals efficiently. Comparing the
performance of Algorithm~\ref{algor3} under the partially noncentered
parametrization with that of the centered and noncentered
parametrizations, we observed that partial noncentering automatically
tends toward the better of centering and noncentering so that it is not
necessary to choose in advance between the centered and noncentered
parametrizations. In many cases, the partially noncentered
parametrization was able to improve upon the fit produced by the better
of centering and noncentering to produce a fit that was closest to that
of MCMC. In terms of computation time, the partially noncentered
parametrization can also provide more rapid convergence when centering
or noncentering is particularly slow. Very often, the lower bound
attained by the partially noncentered parametrization is also higher
than that of the centered and noncentered parametrizations, giving a
tighter lower bound to the log marginal likelihood. To some degree, the
partially noncentered parametrization also alleviates the issue of
underestimation of the posterior variance, leading to some improvement\vadjust{\goodbreak}
in the estimation of the posterior variance, particularly in the fixed
effects which could be centered. Algorithm~\ref{algor3} under the partially
noncentered parametrization thus offers itself as a fast, deterministic
alternative to MCMC methods for fitting GLMMs with improved estimation
compared to the centered and noncentered parametrizations. We also
demonstrate that the variational lower bound produced as part of the
computation in Algorithm~\ref{algor3} can be useful in model
selection.\vspace*{5pt}

\begin{appendix}
\section{Evaluating the Variational Lower Bound}\label{secA}
From (\ref{lowerbound}), (\ref{factorization}) and (\ref{VAGLMM}),
\begin{eqnarray*}
\mathcal{L}& = &\sum_{i=1}^n
S_{y_i} + \sum_{i=1}^n
S_{\tilde{\alpha}_i} + S_\beta+ E_q \bigl\{ \log p(D|\nu,S)
\bigr\} \\
&&{}-E_q\bigl\{ \log q(\beta)\bigr\}-\sum
_i^n E_q\bigl\{ \log q(\tilde{
\alpha}_i)\bigr\}
\\
&&{} - E_q\bigl\{ \log q(D)\bigr\}.
\end{eqnarray*}
To evaluate the terms in the lower bound, we use the following two
lemmas which we state without proof:
%
\begin{lemma}
Suppose $p_1(x)=N(\mu_1,\Sigma_1)$ and\break $p_2(x)=N(\mu_2,\Sigma_2)$ where
$x$ is a $p$-dimensional vector, then
$\int p_2(x)\log p_1(x) \,\mathrm{d}x=-\frac{p}{2}\log(2\pi)-\break{\frac
{1}{2}\log}|\Sigma_1|-\frac{1}{2}(\mu_2-\mu_1)^T\Sigma_1^{-1}(\mu_2-\mu
_1)-\frac{1}{2}\operatorname{tr}(\Sigma_1^{-1}\Sigma_2)$.
\end{lemma}
%
\begin{lemma}
Suppose $p(D)=\mathit{IW}(\nu,S)$ where $D$ is a symmetric, positive definite
$r\times r$ matrix, then
${\int p(D)\log}|D| \,\mathrm{d}D={\log}|S|-\sum_{l=1}^r \psi (\frac{\nu
-l+1}{2} )- r\log2$ and $\int p(D)D^{-1} \,\mathrm{d}D=\nu S^{-1}$
where $\psi(\cdot)$ denotes the digamma function.
\end{lemma}
Using these two lemmas, we can compute most of the terms in
the lower bound:
\begin{eqnarray*}
S_{\beta} &=& \int q(\beta) \log p(\beta|\Sigma_\beta)
\,\mathrm{d}\beta\\
&=&-\frac{p}{2}\log(2\pi)-{\frac{1}{2}\log}|\Sigma_\beta|\\
&&{}- \frac{1}{2}{\mu
_\beta^q}^T\Sigma_\beta^{-1}\mu_\beta^q
-\frac{1}{2}\operatorname{tr}\bigl(\Sigma_\beta^{-1}\Sigma_\beta^q\bigr),
\\
S_{\tilde{\alpha}_i}&=&\int q(\beta)q(D)q(\tilde{\alpha}_i) \log
p(\tilde{\alpha}_i|\beta,D) \,\mathrm{d}\beta \,\mathrm{d}D \,\mathrm
{d}\tilde{\alpha}_i\\
&=&-\frac{r}{2}\log(2\pi)\\
&&{}-\frac{1}{2}  \Biggl\{\log\bigl|S^q\bigr|-\sum_{l=1}^r \psi
\biggl( \frac{\nu^q-l+1}{2}  \biggr)\\
&&\qquad\hspace*{96pt}{} -r\log2 \Biggr\}\\
&&{} -\frac{\nu^q}{2}
\bigl[\bigl(\mu_{\tilde{\alpha}_i}^q-\tilde{W}_i\mu_\beta
^q\bigr)^T {S^q}^{-1}\bigl(\mu_{\tilde{\alpha}_i}^q-\tilde{W}_i\mu_\beta
^q\bigr)\\
&&\hspace*{53.1pt}{}+\operatorname{tr}\bigl\{{S^q}^{-1} \bigl(\Sigma_{\tilde{\alpha}_i}^q+\tilde
{W}_i\Sigma_\beta^q{\tilde{W}_i}^T\bigr)\bigr\} \bigr],\\[-22pt]
\end{eqnarray*}
\begin{eqnarray*}
&&
E_q \bigl\{\log p(D|\nu,S)\bigr\} \\
&&\quad=\int q(D)\log p(D|\nu,S) \,\mathrm{d}D\\
&&\quad=-\frac{\nu^q}{2}\operatorname{tr}\bigl({S^q}^{-1}S\bigr) -\frac{r(r-1)}{4}\log
(\pi)\\
&&\qquad{}-\sum_{l=1}^r \log\Gamma \biggl(\frac{\nu+1-l}{2} \biggr)+{\frac{\nu
}{2}\log}|S|\\
&&\qquad{}-\frac{\nu+r+1}{2}  \Biggl\{\log\bigl|S^q\bigr|-\sum_{l=1}^r \psi \biggl(\frac{\nu
^q-l+1}{2} \biggr)\\
&&\qquad\hspace*{155.5pt}{}-r\log2  \Biggr\}\\
&&\qquad{}-\frac{\nu r}{2}\log2,\\
&&E_q\bigl\{\log q(\beta)\bigr\}\\
&&\quad=\int q(\beta)\log q(\beta) \,\mathrm
{d}\beta\\
&&\quad=-\frac{p}{2}\log(2\pi)-\frac
{1}{2}\log\bigl|\Sigma_\beta^q\bigr|-\frac{p}{2},
\\
&&E_q\bigl\{\log q(\tilde{\alpha}_i)\bigr\}\\
&&\quad=\int q(\tilde{\alpha}_i) \log
q(\tilde{\alpha}_i) \,\mathrm{d}\tilde{\alpha}_i
\\
&&\quad=-\frac{r}{2}\log(2\pi
)-\frac{1}{2}\log\bigl|\Sigma_{\tilde{\alpha}_i}^q\bigr|-\frac{r}{2},
\\
&&E_q\bigl\{\log q(D)\bigr\}\\
&&\quad=\int q(D)\log q(D) \,\mathrm{d}D\\
&&\quad=-\frac{\nu^q r}{2}\log2-\frac{r(r-1)}{4}\log\pi\\
&&\qquad{}-\sum_{l=1}^r \log
\Gamma \biggl(\frac{\nu^q+1-l}{2} \biggr)+\frac{\nu^q}{2} \log\bigl|S^q\bigr|\\
&&\qquad{}
-\frac{\nu^q+r+1}{2}  \Biggl\{ \log\bigl|S^q\bigr|-\sum_{l=1}^r \psi \biggl(\frac{\nu
^q-l+1}{2} \biggr)\\
&&\qquad\hspace*{160pt}{}-r\log2  \Biggr\}\\
&&\qquad{}-\frac{\nu^qr}{2}.
\end{eqnarray*}
The only term left to evaluate is
\[
S_{y_i} = \int q(\beta)q(\tilde{\alpha}_i) \log
p(y_i|\beta,\tilde {\alpha}_i) \,\mathrm{d}\beta
\,\mathrm{d}\tilde{\alpha}_i.
\]
For Poisson responses with the log link function\break [see (\ref{Poisson})],
\begin{eqnarray*}
S_{y_i}&=& y_i^T\bigl\{\log(E_i)+V_i
\mu_\beta^q +X_i^R
\mu_{\tilde{\alpha
}_i}^q\bigr\} -E_i^T
\kappa_i \\
&&{}-1_{n_i}^T \log(y_i!),
\end{eqnarray*}
where $\kappa_i=\exp\{V_i \mu_\beta^q +X_i^R \mu_{\tilde{\alpha
}_i}^q+\frac{1}{2}\operatorname{diag}(V_i \Sigma_\beta^q
{V_i}^T+X_i^R\Sigma_{\tilde{\alpha}_i}^q{X_i^R}^T)\}$. For Bernoulli
responses with the logit link function [see (\ref{logistic})],
\begin{eqnarray*}
S_{y_i}&=& y_i^T\bigl(V_i
\mu_\beta^q +X_i^R
\mu_{\tilde{\alpha}_i}^q\bigr)\\
&&{}-\sum_{j=1}^{n_i}
E_q \bigl[\log\bigl\{1+ \exp\bigl(V_{ij}^T
\beta+{X_{ij}^R}^T \tilde {\alpha}_i
\bigr)\bigr\} \bigr],
\end{eqnarray*}
where $E_q [\log\{1+ \exp(V_{ij}^T \beta+{X_{ij}^R}^T \tilde{\alpha
}_i)\} ]$ is evaluated using adaptive Gauss--Hermite quadrature
(see Appendix~\ref{secB}). The variational lower bound is thus given by
\begin{eqnarray*}
\mathcal{L}&=&\sum_{i=1}^n
S_{y_i}+\frac{1}{2}\sum_{i=1}^n
\log\bigl|\Sigma _{\tilde{\alpha}_i}^q\bigr|\\
&&{}+\frac{1}{2}\log\bigl|
\Sigma_\beta^{-1}\Sigma_\beta ^q\bigr|
-\frac{1}{2}\operatorname{tr}\bigl(\Sigma_\beta^{-1}
\Sigma_\beta^q\bigr)\\
&&{}-\frac
{1}{2}{\mu_\beta^q}^T
\Sigma_\beta^{-1}\mu_\beta^q -
\frac{\nu^q}{2} \log \bigl|S^q\bigr|
\\
&&{} +{\frac{\nu}{2}\log}|S|-\sum_{l=1}^r
\log\Gamma \biggl(\frac{\nu
^q+1-l}{2} \biggr) \\
&&{}+ \sum_{l=1}^r
\log\Gamma \biggl(\frac{\nu+1-l}{2} \biggr)\\
&&{}+\frac
{p+nr}{2}+
\frac{nr}{2}\log2.
\end{eqnarray*}
Note that this expression is valid only after each of the parameter
updates has been made in Algorithm~\ref{algor3}.

\section{Gauss--Hermite Quadrature for Logistic Mixed Models}\label{secB}
We want to evaluate $E_q\{ b(V_{ij}^T \beta+{X_{ij}^R}^T \tilde{\alpha
}_i)\}$ where $b(x)=\log(1+\mathrm{e}^x)$ for each $i=1,\ldots,n$ and
$j=1,\ldots,\break n_i$. Let $\mu_{ij}=V_{ij}^T \mu_\beta^q +{X_{ij}^R}^T
\mu_{\tilde{\alpha }_i}^q$ and $\sigma_{ij}^2=V_{ij}^T \Sigma_\beta^q
V_{ij} +{X_{ij}^R}^T \Sigma_{\tilde{\alpha}_i}^q X_{ij}^R$. Following
\citet{OrmWan12}, we reduce $E_q\{ b(V_{ij}^T \beta+{X_{ij}^R}^T
\tilde{\alpha }_i)\}$ to a univariate integral such that
\begin{eqnarray*}
&&
E_q\bigl\{ b\bigl(V_{ij}^T
\beta+{X_{ij}^R}^T \tilde{\alpha}_i
\bigr)\bigr\}\\
&&\quad=\int_{-\infty
}^{\infty} b(\sigma_{ij}x+
\mu_{ij}) \phi(x;0,1) \,\mathrm{d}x,
\end{eqnarray*}
where $\phi(x;\mu,\sigma)$ denotes the Gaussian density for a random
variable $x$ with mean $\mu$ and standard deviation~$\sigma$. Let
$B^{(r)}(\mu,\sigma)=\int_{-\infty}^{\infty} b^{(r)}(\sigma x+\mu)
\phi(x;0,1) \,\mathrm{d}x$ where $b^{(r)}(x)$ denotes the $r$th derivative of
$b(\cdot)$ with respect to $x$. If $\mu$ and $\sigma$ are vectors, say,
\[
\mu=\left[\matrix{1
\cr
2
\cr
3} \right] \quad\mbox{and}\quad \sigma=\left[\matrix{4
\cr
5
\cr
6} \right],
\]
then
\[
B^{(r)} (\mu,\sigma)=\left[\matrix{ B^{(r)}(1,4)
\vspace*{2pt}\cr
B^{(r)}(2,5)
\vspace*{2pt}\cr
B^{(r)}(3,6)} \right].
\]
For each cluster $i$, let $\mu_i=(\mu_{i1},\ldots,\mu_{in_i})^T= V_{i}
\mu_\beta^q +X_{i}^R \mu_{\tilde{\alpha}_i}^q$ and
\begin{eqnarray*}
\sigma_i&=&(\sigma_{i1},\ldots,\sigma_{in_i})^T
\\[-2pt]
&=&\sqrt{\operatorname{diag}\bigl(V_{i} \Sigma_\beta^q V_{i}^T +X_{i}^R
\Sigma_{\tilde{\alpha}_i}^q {X_{i}^R}^T\bigr)}.
\end{eqnarray*}

We evaluate $B^{(r)}(\mu_{ij},\sigma_{ij})$ using adaptive
Gauss--Hermite quadrature (\cite{LiuPie94}) for each $i=1,\ldots,n$,
$j=1,\ldots,n_i$ and $r=0,1,2$. \citet{OrmWan12} have considered a
similar approach. In Gauss--Hermite quadrature, integrals of the form
$\int_{-\infty}^{\infty} f(x)\mathrm{e}^{-x^2} \,\mathrm{d}x$ are approximated by
$\sum_{k=1}^m w_kf(x_k)$, where $m$ is the number of quadrature points,
the nodes $x_i$ are zeros of the $m$th order Hermite polynomial and
$w_i$ are suitably corresponding weights. This approximation is exact
for polynomials of degree $2m-1$ or less. For low-order quadrature to
be effective, some transformation is usually required so that the
integrand is sampled in a suitable range. Following the procedure
recommended by \citet{LiuPie94}, we rewrite $B^{(r)}(\mu
_{ij},\sigma_{ij})$ as
\begin{eqnarray*}
&&
B^{(r)}(\mu_{ij},\sigma_{ij})\\[-2pt]
&&\quad=\int
_{-\infty}^{\infty} \frac
{b^{(r)}(\sigma_{ij} x+\mu_{ij})\phi(x;0,1)}{\phi(x;\hat{\mu}_{ij},\hat
{\sigma}_{ij})}\phi(x;\hat{
\mu}_{ij},\hat{\sigma}_{ij}) \,\mathrm{d}x
\\[-2pt]
&&\quad= \sqrt{2} \hat{\sigma}_{ij} \int_{-\infty}^{\infty}
\bigl[\mathrm {e}^{x^2} b^{(r)}\bigl(\sigma_{ij}(
\hat{\mu}_{ij}+\sqrt{2}\hat{\sigma }_{ij}x) +
\mu_{ij}\bigr)\\[-2pt]
&&\qquad\hspace*{97.7pt}{}\cdot\phi(\hat{\mu}_{ij}+\sqrt{2}\hat{\sigma
}_{ij}x;0,1) \bigr]\\
&&\qquad\hspace*{48.7pt}{}\cdot \mathrm{e}^{-x^2} \,\mathrm{d}x,
\end{eqnarray*}
which can be approximated using Gauss--Hermite quadrature by
\begin{eqnarray*}
&&
B^{(r)}(\mu_{ij},\sigma_{ij}) \\[-2pt]
&&\quad\approx\sqrt{2} \hat{
\sigma}_{ij} \sum_{k=1}^m
w_k \mathrm{e}^{x_k^2} b^{(r)}\bigl(
\sigma_{ij}(\hat{\mu}_{ij}+\sqrt{2}\hat {\sigma}_{ij}x_k)
+ \mu_{ij}\bigr)\\[-2pt]
&&\hspace*{66.6pt}{}\cdot\phi(\hat{\mu}_{ij}+\sqrt{2}\hat{\sigma
}_{ij}x_k;0,1).
\end{eqnarray*}
For the integrand to be sampled in an appropriate region, we take $\hat
{\mu}_{ij}$ to be the mode of the integrand and $\hat{\sigma}_{ij}$ to
be the standard deviation of the normal density approximating the
integrand at the mode, so that
\begin{eqnarray*}
\hat{\mu}_{ij}&=&\mathop{\arg\max}_x \bigl
\{b^{(r)}(\sigma_{ij}x + \mu_{ij})\phi(x;0,1) \bigr\},
\\[-2pt]
\hat{\sigma}_{ij}&=& \biggl[-\frac{\mathrm{d}^2}{\mathrm{d}x^2}\log \bigl
\{b^{(r)}(\sigma _{ij}x + \mu_{ij})\\[-2pt]
&&\hspace*{75pt}{}\cdot\phi(x;0,1) \bigr
\} \bigg|_{x=\hat{\mu}_{ij}} \biggr]^{-{1/2}}
\end{eqnarray*}
for $j=1,\ldots,n_i$ and $i=1,\ldots,n$. For computational efficiency, we
evaluate $\hat{\mu}_{ij}$ and $\hat{\sigma}_{ij}$, $i=1,\ldots,n$,
$j=1,\ldots,n_i$, for the case $r=1$ only once in each cycle of updates and
use these values for $r=0,2$. No significant loss of accuracy was
observed in doing this. We implement adaptive Gauss--Hermite quadrature
in \texttt{R} using the
\texttt{R} package \texttt{fastGHQuad}
(\cite{Blo}). The quadrature nodes and weights can be obtained via
the function \texttt{gaussHermiteData()} and the
function \texttt{aghQuad()} approximates integrals
using the\break method of \citet{LiuPie94}. We used 10 quadrature points
in all the examples.
\end{appendix}

\section*{Acknowledgments}

Linda S. L. Tan was partially supported as part of Singapore Delft
Water Alliance's tropical reservoir research programme. We thank the
Editors and referees for their constructive comments and suggestions
which have helped to improve the content and clarity of this paper. We
also thank Matt Wand for making available to us his preliminary work on
fully simplified multivariate normal nonconjugate variational message
passing updates and for his careful reading and comments on an earlier
version of this paper.



\end{document}